\documentclass[lettersize,journal]{IEEEtran}
\usepackage[table,xcdraw]{xcolor}
\usepackage{amsmath,amsfonts}
\usepackage[inkscapelatex=false]{svg}
\usepackage{lipsum}
\usepackage{soul}
\usepackage{algpseudocode}
\usepackage{algorithm}
\usepackage{array}
\usepackage[caption=false,font=normalsize,labelfont=sf,textfont=sf]{subfig}
\usepackage{textcomp}
\usepackage{gensymb}
\usepackage{tabularx}
\usepackage[utf8]{inputenc}
\usepackage{booktabs}
\usepackage{stfloats}
\usepackage{svg}
\usepackage{orcidlink} 
\usepackage{float}
\usepackage{url}
\usepackage{verbatim}
\usepackage{multirow}
\usepackage{graphicx}
\usepackage{cite}
\makeatletter
\let\NAT@parse\undefined
\makeatother
\usepackage{hyperref}
\hypersetup{
    colorlinks=true,
    linkcolor=blue,
    citecolor=blue, 
}
\newcommand{\metric}[1]{\fontsize{7pt}{10pt}\selectfont \textbf{#1}}
\definecolor{lightblue}{RGB}{200, 230, 245}
\definecolor{lightpurple}{RGB}{240, 210, 240}
\definecolor{lightgreen}{RGB}{220, 245, 220}
\definecolor{customPurple}{RGB}{215, 110, 204}
\sethlcolor{lightblue}

\renewcommand\footnoterule{
    \kern -3pt
    \hrule width 0.2\columnwidth 
    \kern 2pt} 

\newcommand{\hlblue}[1]{{\sethlcolor{lightblue}\hl{#1}}}
\newcommand{\hlpurple}[1]{{\sethlcolor{lightpurple}\hl{#1}}}
\newcommand{\hlgreen}[1]{{\sethlcolor{lightgreen}\hl{#1}}}

\newcommand{\bmu}{\boldsymbol{\mu}}
\newcommand{\bepsilon}{\boldsymbol{\epsilon}}
\newcommand{\T}{T} 
\definecolor{blue}{rgb}{0,0.3,0.6}

\begin{document}

\title{RN-SDEs: Limited-Angle CT Reconstruction with Residual Null-Space Diffusion Stochastic Differential Equations}

\author{\thanks{This work was supported by National Institutes of Health (NIH) grants U54CA268084, R01CA228272, National Science Foundation (NSF) grant EFMA-1830961 , the Center for Physical Genomics and Engineering at Northwestern University, and philanthropic support from Rob and Kristin Goldman, the Christina Carinato Charitable Foundation, Mark E. Holliday and Mrs. Ingeborg Schneider, and Mr. David Sachs. \textit{(Corresponding author: jiaqi Guo; Aggelos K. Katsaggelos.)}}
Jiaqi Guo$^{\orcidlink{0009-0002-6263-2858}}$, Santiago López-Tapia$^{\orcidlink{0000-0003-2090-7446}}$, Wing Shun Li$^{\orcidlink{0000-0001-9308-3674}}$, Yunan Wu$^{\orcidlink{0000-0001-6980-9746}}$, Marcelo Carignano$^{\orcidlink{0000-0001-8345-7724}}$, Martin Kröger$^{\orcidlink{0000-0003-1402-6714}}$, Vinayak P. Dravid$^{\orcidlink{0000-0002-6007-3063}}$, Igal Szleifer$^{\orcidlink{0000-0002-8708-0335}}$, Vadim Backman$^{\orcidlink{0000-0003-1981-1818}}$ and Aggelos K. Katsaggelos$^{\orcidlink{0000-0003-4554-0070}}$,~\IEEEmembership{Fellow,~IEEE}%
\thanks{Jiaqi Guo, Santiago López-Tapia, Yunan Wu, and Aggelos K. Katsaggelos are with the Department of Electrical and Computer Engineering, Northwestern University, IL 60208 USA (e-mail: jiaqi.guo@northwestern.edu; santiago.lopeztapia@northwestern.edu; yunanwu2020@u.northwestern.edu; a-katsaggelos@northwestern.edu)}%
\thanks{Vinayak P. Dravid is with the Department of Materials Science and Engineering, Northwestern University, Evanston, IL 60208 USA (e-mail: v-dravid@northwestern.edu).}
\thanks{Wing Shun Li is with the Applied Physics Program, Northwestern University, Evanston, IL 60208 USA (e-mail: wingli2025@u.northwestern.edu).}
\thanks{Marcelo Carignano, Igal Szleifer and Vadim Backman are with the Department of Biomedical Engineering, Northwestern University, Evanston, IL 60208 USA (e-mail: marcelo.carignano@northwestern.edu; igalsz@northwestern.edu; v-backman@northwestern.edu).}
\thanks{Martin Kröger is with Magnetism and Interface Physics \& Computational Polymer Physics, Department of Materials, ETH Zurich, Zurich, Switzerland (e-mail: mk@mat.ethz.ch).}
}



\maketitle

\begin{abstract}
Computed tomography is a widely used imaging modality with applications ranging from medical imaging to material analysis. One major challenge arises from the lack of scanning information at certain angles, resulting in distortion or artifacts in the reconstructed images. This is referred to as the Limited Angle Computed Tomography (LACT) reconstruction problem. To address this problem, we propose the use of Residual Null-Space Diffusion Stochastic Differential Equations (RN-SDEs), which are a variant of diffusion models that characterize the diffusion process with mean-reverting (MR) stochastic differential equations. To demonstrate the generalizability of RN-SDEs, we conducted experiments with two different LACT datasets, ChromSTEM and C4KC-KiTS. Through extensive experiments, we demonstrate that by leveraging learned MR-SDEs as a prior and emphasizing data consistency using Range-Null Space Decomposition (RNSD) based rectification, we can recover high-quality images from severely degraded ones and achieve state-of-the-art performance in most LACT tasks. Additionally, we present a quantitative comparison of RN-SDE with other networks, in terms of computational complexity and runtime efficiency, highlighting the superior effectiveness of our proposed approach.\footnote{The code will become available upon acceptance of the paper}
\end{abstract}

\begin{IEEEkeywords}
    Limited-angle CT reconstruction, Diffusion stochastic differential equations, Mean-reverting diffusion, Range-null space decomposition
\end{IEEEkeywords}

\section{Introduction}
\label{sec:intro}

\begin{figure}[htbp]
  \centering
  \includegraphics[width=0.5\textwidth]{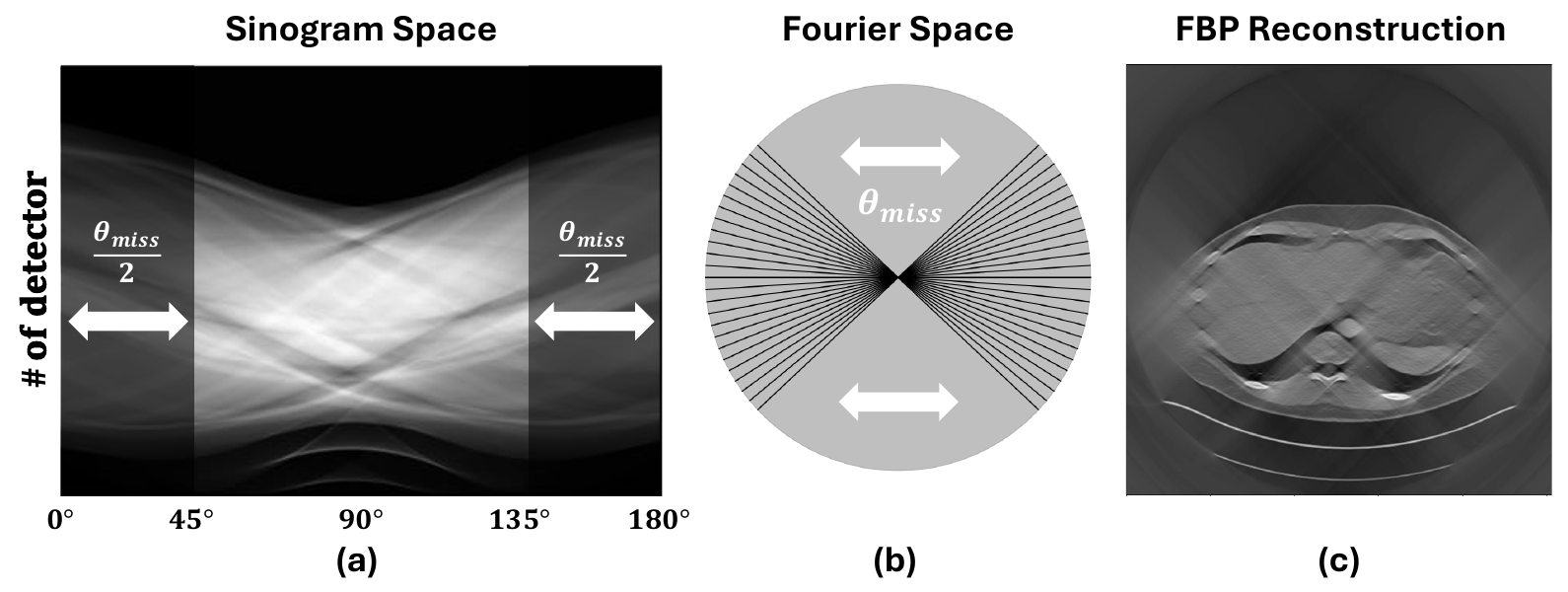}
  \caption{Illustration of limited angle tomography, with missing angles, $\theta_{\rm miss}$, set to be 90\degree. (a) Sinogram of an exemplary 2D observation with certain angles missing; (b) Sampling process in the Fourier space; (c) Reconstruction using the FBP algorithm, with distortion and artifacts present.}
  \label{fig:demoLACT}
\end{figure}

Tomographic reconstruction is a method to interpret an object from its lower dimensional projections. It is widely used in many imaging applications such as X-ray computed tomography, electron tomography and positron emission tomography~\cite{withers2021x, ercius2015electron, muehllehner2006positron}. The forward process can be mathematically represented by the Radon Transform~\cite{radon69u}, in which a series of line integrals are calculated for each projection angle, representing the cumulative information along certain scanning trajectories. Ideally, projections obtained from a full range dense set of projection angles would contain the entire Fourier domain information of the observed object, allowing a direct and precise reconstruction. However, in situations such as electron tomography, experimental setups can hinder uniform projection sampling, restricting information to limited angles. This issue poses a challenge to conventional reconstruction methods such as Filtered Back Projection (FBP)~\cite{dudgeon1983multidimensional} and is referred to as the \textbf{Limited Angle Computed Tomography (LACT)} problem. The Fourier slice theorem~\cite{bracewell1956strip} states that the Fourier transform of the one-dimensional projection of an object corresponds to a line through the center of its two-dimensional (2D) Fourier transform. We pictorially illustrate the LACT problem in Fig.~\ref{fig:demoLACT} and denote the missing angles as $\theta_{\rm miss}$.

\noindent\textbf{LACT Restoration}: Consider the following degradation model,
\begin{equation}\label{inverse_problem}
    \mathbf{y}=\mathbf{A}\mathbf{x}+\mathbf{n},
\end{equation}
\IEEEpubidadjcol where $\mathbf{x}\in\mathbb{R}^{M}$ represent the original image, $\mathbf{y}\in \mathbb{R}^{N}$ its degraded version, and $\mathbf{n}$ is additive white Gaussian noise. The exact form of the degradation operator $\mathbf{A}\in \mathbb{R}^{N\times M}$ depends on the application. In the context of LACT, $\mathbf{A}$ denotes the Radon transform that defines a mapping from an image to a set of projections, which can be presented as a sinogram. Sinogram describes the unique pattern formed by the Radon transform of an object, resembling a series of sine waves. The image restoration (IR) problem is the recovery of $\mathbf{x}$ given the observation $\mathbf{y}$ and knowledge of $\mathbf{A}$. A typical solution to the problem in Eq.~\ref{inverse_problem} is the result of the following optimization:
\begin{equation}
\label{eq2}
\hat{\mathbf{x}}=\underset{\mathbf{x}}{\arg \min } \frac{1}{2 \sigma^2}\|\mathbf{A} \mathbf{x}-\mathbf{y}\|_2^2+\lambda \mathcal{R}(\mathbf{x}) .
\end{equation}
The term $\frac{1}{2 \sigma^2}\|\mathbf{A} \mathbf{x}-\mathbf{y}\|_2^2$ ensures the fidelity to the data, whereas the regularization term, $\mathcal{R}(\mathbf{x})$, incorporates prior knowledge about the solution into the restoration process, e.g., Tikhonov regularization, total variation distance, transform-domain sparsity and dictionary learning ~\cite{alberti2021learning, barutcu2021limited, danielyan2011bm3d, kudo2013image}, while the regularization parameter $\lambda$ controls the relative contribution of the two terms.

Deep learning (DL) represents a recent tool for solving LACT tasks. One straightforward application of DL is to train a convolutional neural network (CNN), $\mathcal{Z}(\cdot)$, to directly perform a regularized inversion of the Radon transform~\cite{anirudh2018lose, gupta2018cnn, han2018framing, jin2017deep, zhang2020artifact, zhang2021cd}. Typically, one seeks to minimize the mean squared error (MMSE) between the restored image $\mathcal{Z}(\mathbf{y})$, and the ground truth (GT) images $\mathbf{x}$. These MMSE-based methods circumvent the explicit modeling of $\mathbf{A}$ and the data priors, allowing for fast inference. However, in cases of acute degradation where information is irreversibly lost, MMSE-based methods tend to produce washed-out reconstructions that lack high-frequency details. Essentially, the MMSE solution can be thought of as the average of all potential solutions, i.e., the conditional mean of the posterior distribution $\mathbb{E}[\mathbf{x} | \mathbf{y}]$~\cite{kawar2021snips}.

The advent of generative models, notably Generative Adversarial Networks (GANs)~\cite{goodfellow2014generative}, Variational Autoencoders (VAEs)~\cite{kingma2013auto} and diffusion models~\cite{ho2020denoising}, along with their popular variants~\cite{van2017neural, karras2019style, song2019generative, song2020score}, have greatly increased the performance of DL-based IR methods. Their strength lies in their ability to produce perceptually realistic results from a meaningful latent space. Similarly to Eq.~\ref{eq2}, given a pre-trained generative model $\mathcal{G}(\cdot)$, the optimization objective can be formulated as,
\begin{equation}\label{eq4}
\underset{\mathbf{w}}{\arg \min } \frac{1}{2 \sigma^2}\|\mathbf{A} \mathcal{G}(\mathbf{w})-\mathbf{y}\|_2^2+\lambda \mathcal{R}(\mathbf{w}),
\end{equation}
where $\mathbf{w}$ represents the latent variables, $\mathcal{G}(\mathbf{w})$ represents the generated samples conditioned on $\mathbf{w}$, and $\mathcal{R}(\mathbf{w})$ represents the data prior that regularizes $\mathbf{w}$ to its original distribution space. 
\\
\begin{figure}[tbp]
    \centering
    \includegraphics[width=0.485\textwidth]{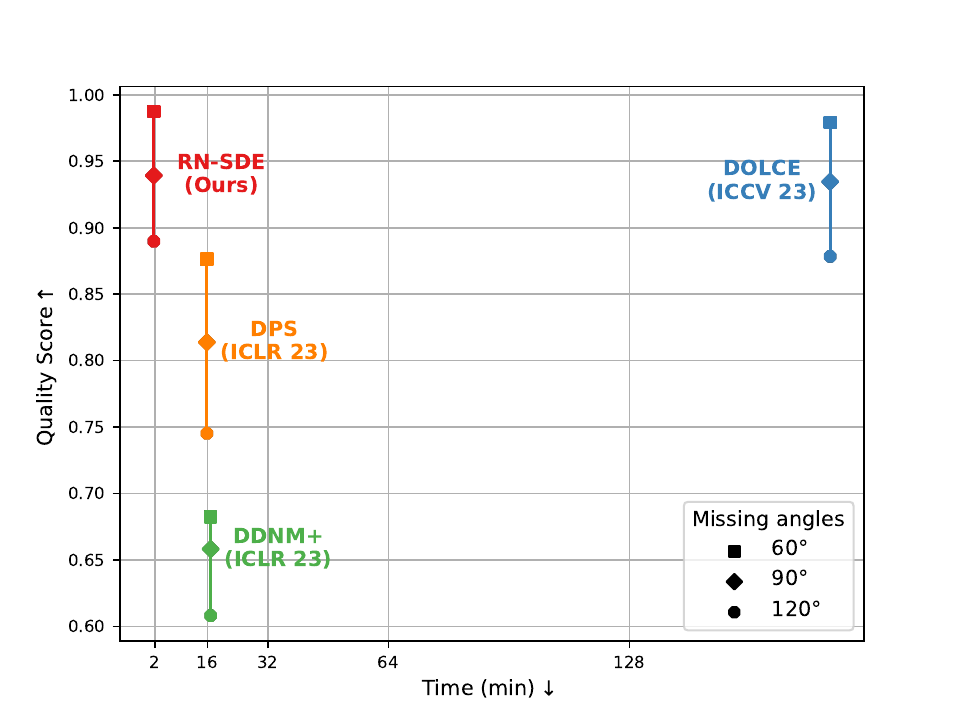}
    \caption{Quantitative performance evaluation on the C4KC-KiTS dataset~\cite{heller2019kits19}. This plot demonstrates the relationship between image quality and average processing time per reconstruction in minutes. The quality is measured as the average value of the relative performance of each method in each of the three metrics (PSNR, SSIM, and LPIPS~\cite{zhang2018unreasonable}). Our proposed RN-SDE method obtains, on average, higher quality while having significantly lower computational complexity. More details are provided in Section~\ref{experiment: compare}.}
    \label{fig:SOTA}
\end{figure}

\noindent\textbf{Diffusion Models and their Limitations:} Diffusion models have recently outperformed GANs and emerged as the new state-of-the-art (SOTA) generative models~\cite{dhariwal2021diffusion}. To start with, most current studies on diffusion models are primarily focused on three main formulations~\cite{croitoru2023diffusion, yang2023diffusion}: denoising diffusion probabilistic models (DDPMs)~\cite{ho2020denoising, nichol2021improved, sohl2015deep}, score-based generative models (SGMs)~\cite{song2019generative, song2020improved}, and stochastic differential equations (SDEs) ~\cite{song2021maximum, song2020score} based models. Among them, DDPMs and SGMs  can be described using SDEs with infinite time steps or noise levels. Importantly, all formulations mentioned above are trained to model the Markov transition from a simple distribution, e.g., Gaussian distribution, to a complex data distribution. This process is facilitated by a learned score function~\cite{liu2016kernelized, chwialkowski2016kernel} that assesses the likelihood of intermediate transition probability distributions, enabling the generation of samples through a series of controlled stochastic processes. Unlike the MMSE-based method, the solutions of diffusion can be considered as sampling from the posterior distribution $p(\mathbf{x}|\mathbf{y})$, thus resulting in better perceptual quality. Recognizing the potential, several studies have started to exploit the capability of diffusion models to address different IR tasks~\cite{wang2023zero, choi2021ilvr, chung2023diffusion, chung2022improving} as well as the LACT problem~\cite{song2021solving, liu2023dolce}. To do this, most of these methods keep the training of a diffusion model intact and only modify the inference procedure to generate samples that satisfy realness and fidelity to the data. In other words, the pre-trained diffusion model will be capable of addressing any linear inverse problem within the same image space. 

It is widely recognized that the reverse sampling process of the diffusion models is often time-consuming due to its dependence on an iterative Markov Chain Monte Carlo (MCMC) process, which typically requires a few thousand steps to converge. A potential solution to accelerate the sampling could involve taking larger steps toward the desired data distribution at each iteration. However, this approach could potentially derail the sampling process from the beginning due to the inaccurate estimation of the score function in areas of low density. Luo et al.~\cite{luo2023image} proposed a mean-reverting SDEs (MR-SDEs) approach that initiates the reverse diffusion process from a Gaussian distribution, with the mean set equal to the observed measurement. This method can significantly reduce the number of iterations required for convergence. However, one remaining question is whether the constraints imposed by the degraded measurements are strong enough to produce proper reconstructions. Based on our experiment, the answer is generally negative. Despite our implementation of mean-reverting strategies to narrow down the possible solutions, numerous candidate HQ images could still explain the observed degraded measurements, particularly in cases of severe degradation.

Therefore, an additional consistency constraint will be needed to address the above issue, highlighting another challenge associated with the diffusion model, \textit{i.e.,} the perceptual-distortion tradeoff~\cite{blau2018perception}. Several studies have revealed that while images restored by diffusion models exhibit superior perceptual quality, they may not perform well with respect to fidelity metrics, such as SSIM and PSNR~\cite{saharia2022image, li2022srdiff, kawar2021snips}. The range null space decomposition (RNSD)~\cite{schwab2019deep, wang2023zero} presents a novel solution for balancing realness with consistency. This method leverages the diffusion model to produce realistic content in the null space and analytically ensures data consistency in the range space. We believe RNSD is an ideal tool to guide the reverse diffusion of MR-SDEs.

\noindent\textbf{Our Proposed Work}: Based on the discussion above, in this paper, we propose a novel approach to combine RNSD with MR-SDE for improved LACT reconstruction. Given a limited-angle sinogram, we first employ a standard method (any end-to-end DNN or an analytic CT reconstruction algorithm) to obtain a seed reconstruction, which is then utilized to condition and initialize the mean-reverting reverse diffusion process. During inference, we introduce an RNSD-based approach to enforce data consistency. This approach directly approximates the Radon pseudo-inverse~\cite{penrose1955generalized} without computing the matrix-form operator. More importantly, its training objective can be easily generalized to other IR problem settings, broadening the applicability of our proposed algorithm. Finally, since the actual generated content of this diffusion process is the Residual/Difference between the restored and degraded image pairs, we refer to our proposed method as the \textit{\textbf{R}esidual \textbf{N}ull Space Diffusion \textbf{S}tochastic \textbf{D}ifferential \textbf{E}quations} (RN-SDEs) method. To summarize, our main contributions are as follows:
\begin{itemize}
    \item We propose RN-SDEs, a diffusion model that uses NafNet~\cite{chen2022simple} as the diffusion backbone, combining \textit{mean-reverting SDEs (MR-SDEs)}~\cite{luo2023image} and \textit{the range null space decomposition (RNSD)}, for fast and high-quality LACT reconstruction.    
    \item We utilize the MMSE solution, $\mathbb{E}[\mathbf{x}|\mathbf{y}]$, specifically the NafNet reconstructions, as the "mean" in MR-SDEs, significantly enhancing the fidelity of the reconstructions.
    \item We introduce a universal approach to approximate the pseudo-inverse operation without accessing the operator $\mathbf{A}^{\dagger}$, and improve the application of data consistency in~\cite{wang2023zero}.
    \item We demonstrate the effectiveness of RN-SDEs with ChromSTEM and medical CT datasets. Experimental results indicate that we can yield superior perceptual quality while preserving data consistency. Moreover, as can be seen in Fig.~\ref{fig:SOTA}, RN-SDEs are fast during inference, significantly outpacing DOLCE~\cite{liu2023dolce}, the SOTA for the solution of the LACT problem, by hundreds of times, while producing a better quality output. 
\end{itemize}

\section{Preliminaries}
\label{sec:Preliminaries}

\subsection{Diffusion Stochastic Differential Equations (SDEs)} \label{sdes}
Diffusion SDEs~\cite{song2020score} provide a mathematical way to formulate both the forward and reverse processes in a diffusion model as the solutions of stochastic differential equations. Specifically, the forward process $\{\mathbf{x}_t\}_{t=0}^\T$ involves perturbing clean data samples $\mathbf{x}_0\sim{p(\mathbf{x}_0})$ to a noisy version $\mathbf{x}_\T\sim{p(\mathbf{x}_\T})$ governed by the following SDE~\cite{song2020score},
\begin{equation}\label{eq5}
\mathrm{d} \mathbf{x}=\mathbf{f}(\mathbf{x}, t) \mathrm{d} t+g(t) \mathrm{d} \mathbf{w},
\end{equation}
where $\textbf{w}$ denotes a standard Wiener process, $\mathbf{f}(\cdot)$ is a function of $\mathbf{\mathbf{x}}$ and $t$ that computes the drift coefficient, and $g(\cdot)$ is a time-dependent function that computes the diffusion coefficient. Normally, the final state $\mathbf{x}_\T$ is expected to follow a Gaussian distribution characterized by a predefined mean and variance. Note that DDPMs~\cite{ho2020denoising} and SGMs~\cite{song2019generative} can take the form of Eq.~\ref{eq5} by replacing the drift and diffusion coefficients with specific functions as shown in Appendix~\ref{Appendix: DDPMs SGMs to SDEs}. 

To reverse the above process and sample data from noise, Anderson~\cite{elliott1985reverse} shows that a corresponding reverse version exists for any forward SDE defined in the form of Eq.~\ref{eq5}. This reverse SDE runs in the reverse direction but tracks the same marginal distribution $p(\mathbf{x}_t)$, and can be written as,
\begin{equation}\label{eq: reverse SDE}
\mathrm{d} \mathbf{x}=\left[\mathbf{f}(\mathbf{x}, t)-g(t)^2 \nabla_{\mathbf{x}} \log p(\mathbf{x}_t)\right] \mathrm{d} t+g(t) \mathrm{d} \widehat{{\mathbf{w}}},
\end{equation}
where $\widehat{\mathbf{w}}$ is a reverse-time Wiener process and $\nabla_{\mathbf{x}} \log p(\mathbf{x}_t)$ is the score function, which indicates the direction that maximizes the likelihood of the distribution. Typically, the score function is estimated by a neural network $\mathbf{s}_\phi(\mathbf{x}_t, t)$ and trained using the following score matching~\cite{hyvarinen2005estimation, vincent2011connection, song2020sliced} objectives which can be directly estimated from the training data $\mathbf{x}_0$:
\begin{equation}\label{denoising score matching}
\mathbb{E}_{t \in \mathcal{U}}  \mathbb{E}_{p(\mathbf{x}_t)}\left[\left\|\nabla_{\mathbf{x}} \log p(\mathbf{x}_t|\mathbf{x}_0)-\mathbf{s}_\phi(\mathbf{x}_t, t)\right\|\right],
\end{equation}
where $\mathcal{U}$ denotes a uniform distribution over the time interval $[0, \T]$. Once trained, the learned score function can be plugged into Eq.~\ref{eq: reverse SDE} to simulate the reverse stochastic process from $\mathbf{x}_\T$ back to $\mathbf{x}_0$. This simulation is carried out using numerical SDE solvers, such as the Euler-Maruyama, Milstein, and stochastic Runge-Kutta methods.

\subsection{Range Null Space Decomposition (RNSD) via \texorpdfstring{$\mathbf{A}^{\dagger}$}{Pseudo-Inverse}} \label{RNSD}
For a linear operator $\mathbf{A}\in\mathbb{R}^{N\times M}$, its pseudo-inverse $\mathbf{A}^{\dagger} \in \mathbb{R}^{M \times N}$ satisfies the Moore-Penrose conditions, with $\mathbf{A}\mathbf{A}^{\dagger}\mathbf{A} \equiv \mathbf{A}$ being one of the important ones for our discussion. 
\begin{itemize}
    \item \noindent\textbf{Range Space Projection:} The product $\mathbf{A}^{\dagger}\mathbf{A}$ acts as a range space projector, mapping any vector $\mathbf{x}\in\mathbb{R}^M$ onto the range space of $\mathbf{A}$, ensuring it corresponds to a vector in $\mathbf{A}$'s column space.
    
    \item \noindent\textbf{Null Space Projection:} The matrix $\mathbf{I} - \mathbf{A}^{\dagger}\mathbf{A}$ serves as a null-space projector, mapping any vector $\mathbf{x}$ onto the corresponding null space of $\mathbf{A}$. This guarantees that the projection, $(\mathbf{I} - \mathbf{A}^{\dagger}\mathbf{A})\mathbf{x}$, is orthogonal to the range space.
    
    \item \noindent\textbf{IR Perspective:} Given an image $\mathbf{x}$, the range space of $\mathbf{A}$ contains the uncorrupted information that must be preserved during restoration. Conversely, the null-space content highlights the information orthogonal to $\mathbf{A}$'s range space, affected by the degradation operation $ \mathbf{A} $. These orthogonal components emerge as potential candidates for correction or refinement during restoration. Noteworthy, by applying $\mathbf{A}$ to the range/null space content, the range space content will become $\mathbf{y}$ (\textit{i.e.,} $ \mathbf{A} \mathbf {\mathbf{A}}^{\dagger} \mathbf{\mathbf{A}} \mathbf{x}\equiv \mathbf{y} $ ) while the null-space content will become $0$ (\textit{i.e.,} $\mathbf{A}(\mathbf{I} - \mathbf{A}^{\dagger}\mathbf{A})\mathbf{x}\equiv \mathbf{0}$).
\end{itemize}
Leveraging these properties, any vector $\mathbf{x}$ can be represented as a composite of its components in the range/null spaces. This leads to:
\begin{equation}\label{eq: decomposition}
    \mathbf{x} \equiv \underbrace{\mathbf{\mathbf{A}}^{\dagger}\mathbf{\mathbf{A}}\mathbf{x}}_{\text{range space}} + \underbrace{(\mathbf{I} - \mathbf{\mathbf{A}}^{\dagger}\mathbf{\mathbf{A}})\mathbf{x}}_{\text{null space}},
\end{equation}
which reveals the inherent partition of the vector space induced by $\mathbf{\mathbf{A}}$. This simple decomposition is significant in analyzing linear inverse problems and will be further explored in subsequent sections.

\section{Method} 
\label{sec:method}
\subsection{LACT Range-Null Space Decomposition} 
In this paper, we will consider the simpler, noise-free scenario of LACT Reconstruction, that is,
\begin{equation}
    \mathbf{y}=\mathbf{A}\mathbf{x},
\end{equation}
where $ \mathbf{x} \sim p(\mathbf{x})$ is an image sampled from the ground-truth (GT) data distribution, $\mathbf{A}$ denotes the Radon transform, and $\mathbf{y}$ is the limited-angle sinogram, and $0 \leq \theta \leq \theta_{\max} < \pi$. Referring to Eq.~\ref{eq4}, our objective is to yield an estimate, $ \hat{\mathbf{x}}$, that satisfies both the consistency constraint, $\mathbf{A} \hat{\mathbf{x}} \approx \mathbf{y}$, and the realness constraints,  $\hat{\mathbf{x}} \sim p(\mathbf{x})$. Among them, the consistency constraint can be analytically guaranteed within the range-space of $\mathbf{A}$ by solving: 
\begin{equation}
\mathbf{A}^{\dagger}\mathbf{y}=\mathbf{A}^{\dagger}\mathbf{A}\hat{\mathbf{x}}.
\end{equation}
The solution $\hat{\mathbf{x}}$ is not unique; it only ensures that the sinogram of the reconstructed image matches the GT image for the available angles, $\theta \in [0, \theta_{\max}]$. More importantly, the consistency constraint is only related to the range space. We can therefore utilize the diffusion models to generate suitable null-space content, $(\mathbf{I} - \mathbf{\mathbf{A}}^{\dagger}\mathbf{\mathbf{A}})\hat{\mathbf{x}}$, that allows the final solutions, $\hat{\mathbf{x}} = \mathbf{\mathbf{A}}^{\dagger}\mathbf{y} + (\mathbf{I} - \mathbf{\mathbf{A}}^{\dagger}\mathbf{\mathbf{A}})\hat{\mathbf{x}}$, to fulfill the realness constraint.

\subsection{Residual Null-Space Diffusion Stochastic Differential Equation (RN-SDE)}\label{RN-SDE}
\noindent\textbf{Forward MR-SDE:} Luo et al.~\cite{luo2023image}, constructed the MR-SDEs as a special case of the SDEs discussed in Sec.~\ref{sdes}. To make the mean \textit{revertible}, they formulate the forward SDE~\cite{luo2023image} as, 
\begin{equation}\label{eq: fw SDE}
\mathrm{d} \mathbf{x}=\zeta_t(\bmu-\mathbf{x}) \mathrm{d} t+\sigma_t \mathrm{d} \mathbf{w},
\end{equation}
where $\zeta_t, \sigma_t$ are time-dependent parameters that respectively regulate the pace of mean-reversion and the noise perturbation and $\bmu\in \mathbb{R}^{M}$ is the mean of the terminal state, which, in the context of IR, usually indicates the degraded images. Note that we have slightly adjusted the symbols here to avoid repetition. In the LACT problem, directly setting the sinogram as $\bmu$ is infeasible due to the dimensionality discrepancy between sinograms and images. A typical workaround~\cite{jin2017deep, liu2023dolce} is to replace the sinogram with a low-fidelity reconstruction, derived from standard inversion methods, such as FBP. Specifically, we incorporate an additional NafNet~\cite{chen2022simple} block for post-processing. This model is optimized with the MMSE objective and the outcomes will be used as $\bmu$ to condition the reverse diffusion process. We demonstrate in Sec.~\ref{sec: experiments} that this novel preprocessing step effectively improve the reconstruction fidelity leveraging the "smoothed" information from MMSE solutions.

Following the setting in~\cite{luo2023image}, we define $\bar{\zeta}_{t}:=\int_0^t \zeta_z \mathrm{~d} z$ and set $\sigma_t^2 / \zeta_t=2 \lambda^2$ for all times $t$, where $\lambda^2$ is the fixed variance. Given the starting state $\mathbf{x}_0$, we can sample $\mathbf{x}_t$ at any arbitrary time $t$ from the following Gaussian distribution:
\begin{equation}\label{eq: MR-SDE}
    p(\mathbf{x}_t | \mathbf{x}_0) = \mathcal{N}(\mathbf{x}_t | \mathbf{m}_t(\mathbf{x}_0), v_t),
\end{equation}
where the mean $\mathbf{m}_t(\mathbf{x}_0)$ and variance $v_t$ are respectively:
\begin{equation}
 \mathbf{m}_t(\mathbf{x}_0):= \bmu + (\mathbf{x}_0 - \bmu) e^{-\bar{\zeta}_{t}},
\end{equation}
\begin{equation}
    v_t := \lambda^2\left(1 - e^{-2 \bar{\zeta}_{t}}\right).
\end{equation}
As noted in~\cite{luo2023image}, Equation~\ref{eq: MR-SDE} is described as \textit{mean-reverting} because, as $t \to \infty$, we will have: $\lim_{t \to \infty} \mathbf{m}_t = \bmu$, and $\lim_{t \to \infty} v_t = \lambda^2$. Such that, the terminal distribution $p(\mathbf{x}_\T)$ can be approximated by a noisy version of the degraded images $\bmu$. 
\\

\noindent\textbf{Reversing the MR-SDE:} To recover the target reconstructed image from $\mathbf{x}_\T$, Luo et al.~\cite{luo2023image} define the reverse MR-SDE as,
\begin{equation}\label{reverse mr sde}
\mathrm{d} \mathbf{x}=\left[\zeta_t(\bmu-\mathbf{x})-\sigma_t^2 \nabla_\mathbf{\mathbf{x}} \log p(\mathbf{x}_t)\right] \mathrm{d} t+\sigma_t \mathrm{d} \widehat{\mathbf{w}},
\end{equation}
where the $\nabla_\mathbf{x} \log p(\mathbf{x}_t)$ is the score of the marginal distribution $p(\mathbf{x}_t)$. Typically, this score function can be estimated using the score-matching objective mentioned in Sec.~\ref{sdes}. However, as stated in~\cite{luo2023image}, this simple objective might lead to an unstable training process. Therefore, they propose an alternative approach to derive an optimum reverse state $\mathbf{x}_{t-1}$ that maximizes the likelihood of $p(\mathbf{x}_{t-1} | \mathbf{x}_t, \mathbf{x}_0)$. Significantly, we found out that the modified objective is equivalent to estimating the following score:
\begin{equation}\label{opt score}
\nabla_\mathbf{x} \log p(\mathbf{x}_t | \mathbf{x}_0)^* =  \frac{g_t\mathbf{x}_t+h_t\mathbf{x}_0-(h_t+g_t)\bmu}{\sigma_t^2 \mathrm{d} t}. 
\end{equation}
Here, by letting $\zeta_t^{\prime}:=\int_{t-1}^t \zeta_z \mathrm{d} z$, the terms $g_t, h_t$ are respectively equal to,
\begin{equation}
\begin{aligned}
& g_t =  \frac{1-e^{-2\bar{\zeta}_{t-1}}}{1-e^{-2 \bar{\zeta}_t}} e^{-\zeta_t^{\prime}}-\zeta_t \mathrm{d} t-1,\\
& h_t = \frac{1-e^{-2 \zeta_t^{\prime}}}{1-e^{-2 \bar{\zeta}_t}} e^{-\bar{\zeta}_{t-1}}.\\
\end{aligned}
\end{equation}
The derivation is detailed in Appendix~\ref{Appendix: RNSD for RN-SDE}. To approximate this score, we propose to use the Conditional NafNet (Cond-NafNet)~\cite{luo2023refusion}, $\mathbf{s}_\phi(\mathbf{x}_t, \bmu, t)$, as the diffusion backbone. This network takes the state $\mathbf{x}_t$, the low-fidelity reconstruction $\bmu$, and time $t$ as input, and is trained with the commonly-used objective described in Eq.~\ref{denoising score matching}.\\

\begin{figure*}[htbp]
  \centering
  \includegraphics[width=0.98\textwidth]{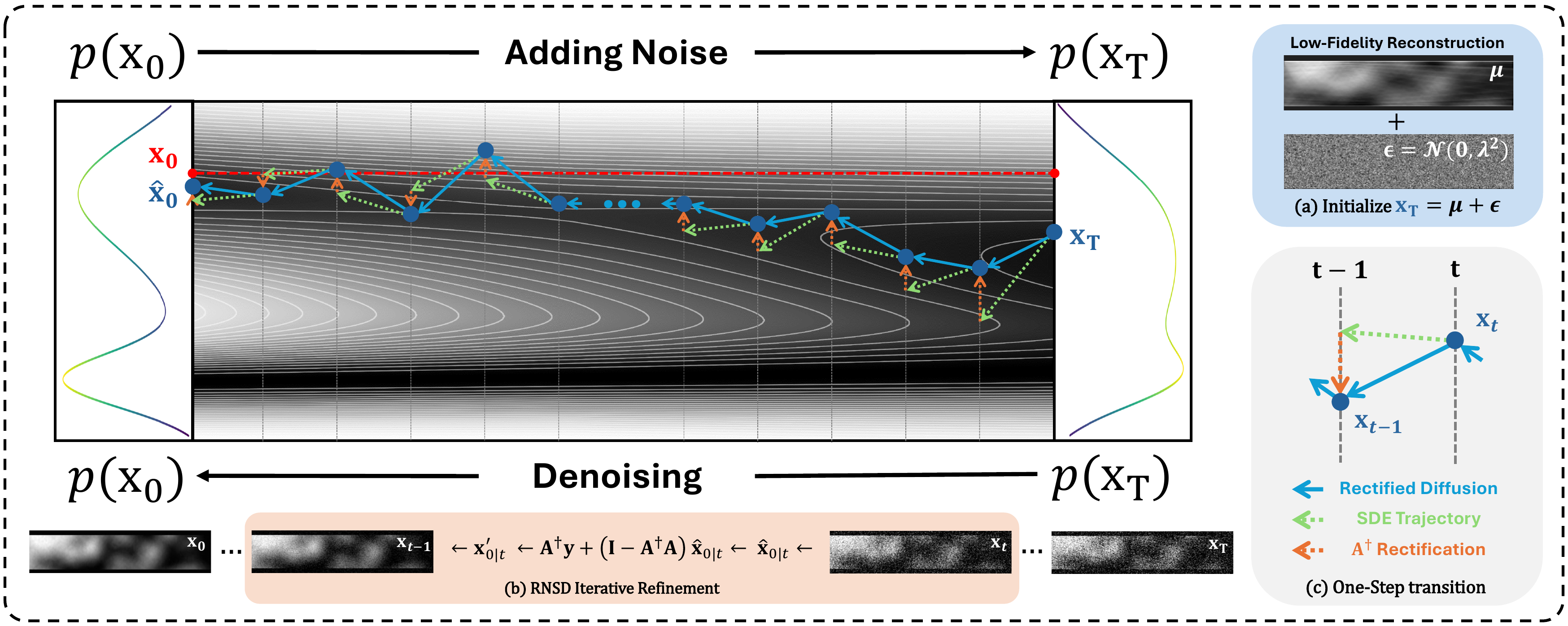}
  \caption{Visualization of our RN-SDE denoising diffusion process for LACT reconstruction. (a) Initialize the terminal state $\hat{\mathbf{x}}_\T$ of diffusion as the summation of a low-quality reconstruction $\bmu$ and a Gaussian noise $\bepsilon$. This step positions the starting point of the reverse diffusion closer to $\mathbf{x}_0$
  within the distribution space, facilitating a more effective and accurate convergence towards the GT image during the denoising stage; (b) illustrates the rectification mechanism involved in the reverse diffusion process, where we applied range-null space decomposition (RNSD) to the intermediate clean prediction $\hat{\mathbf{x}}_{0|t}$ to enforce data consistency; (c) shows the dynamics of a single transition step in the denoising process. Providing a visual example of how the RNSD-based rectification effectively reduces stochasticity during the reverse diffusion process.}
  \label{fig:diffusion}
\end{figure*}

\noindent\textbf{RNSD-Based Iterative Refinement:} During inference, we iteratively simulate the reverse stochastic process from $\mathbf{x}_\T$ back to $\mathbf{x}_0$. In the distribution space, the mean-reverting strategy allows us to initialize $\mathbf{x}_\T$ from a position closer to the GT solution. This leads to a situation where reverse diffusion is more likely to converge close to the GT solution despite the randomness of the inference process. Nevertheless, when the distribution $p(\bmu)$, differs significantly from $p(\mathbf{x}_0)$, i.e., in certain LACT tasks where a large number of angles are missing, the stochasticity of the reverse diffusion process can also possibly harm the consistency of the data. To address this issue, similarly to~\cite{wang2023zero}, we propose to enforce data consistency directly through the RNSD. This is achieved by incorporating an additional rectification into the estimated score at each time step. To be more precise, we first utilize the estimated score, denoted by $\hat{\mathbf{s}}_\phi(\mathbf{x}_t, \bmu, t)$, to generate a clean intermediate state $\hat{\mathbf{x}}_{0|t}$, which is given by,
\begin{equation}\label{clean inter-states}
  \hat{\mathbf{x}}_{0|t} = - \frac{g_t}{h_t}\mathbf{x}_t + \frac{\sigma_t^2}{h_t} \hat{\mathbf{s}}_\phi(\mathbf{x}_t, \bmu, t)  \mathrm{d} t + \left(1+\frac{g_t}{h_t}\right)\bmu.
\end{equation}
This intermediate state can be interpreted as a rough estimation of $\mathbf{x}_0$ at time $t$. Later, we decompose $\hat{\mathbf{x}}_{0|t}$ into $\mathbf{A}$'s range and null space. To promote data consistency, we replace the range-space content by $\mathbf{A}^{\dagger}\mathbf{y}$, while preserving the null-space content, $(\mathbf{I}-\mathbf{A}^{\dagger}\mathbf{A})\hat{\mathbf{x}}_{0|t}$. In such a way, the data consistency can be strictly enforced by the combined solution ${\mathbf{x}}_{0|t}^{\prime}$, which is given by:
\begin{equation}\label{rectification}
    {\mathbf{x}}_{0|t}^{\prime} = \mathbf{A}^{\dagger}\mathbf{y} + (\mathbf{I} - \mathbf{A}^{\dagger}\mathbf{A})\hat{\mathbf{x}}_{0|t}. 
\end{equation}
Finally, the rectified clean estimate ${\mathbf{x}}_{0|t}^{\prime}$ will be inserted into the reverse MR-SDE to sample the next state $\mathbf{x}_{t-1}$ from the modified conditional distribution $p(\mathbf{x}_{t-1}| \mathbf{x}_t, {\mathbf{x}}_{0|t}^{\prime})$. This process can be described using the reverse RN-SDE, which we defined as,
\begin{equation}\label{xt-1}
\begin{aligned}
  \mathrm{d}\mathbf{x} =  (g_t + \zeta_t\mathrm{d}t)&(\bmu-\mathbf{x}_t) \\
   + h_t&(\bmu-{\mathbf{x}}_{0| t}^{\prime}) + \sigma_t \mathrm{d} \widehat{\mathbf{w}}.
\end{aligned}
\end{equation}
Importantly, since the realness is already guaranteed through the maximum likelihood objective during training, by iterating through Eqs.~\ref{clean inter-states},~\ref{rectification} and~\ref{xt-1} from $\T$ down to $1$, we obtain a clean reconstruction that satisfies both the consistency and realness constraints. \\

\noindent\textbf{Approximate the Radon Pseudo-Inverse Operation:} A remaining issue is that the Radon pseudo-inverse operator in Eq.~\ref{rectification} remains unknown to us. Conventional approaches for computing it are typically impractical, primarily due to the computation complexity involved in large-scale matrix operations, which we detailed in Appendix~\ref{Appendix: complexity of A+}. To handle this issue, rather than computing $\mathbf{A}^{\dagger}\in \mathbb{R}^{M\times N}$, we proposed to approximate this operation directly using a neural network ${\mathbf{A}}_\eta^{\dagger}(\cdot)$, with learnable parameters $\eta$.
\begin{figure*}[htb]
  \centering
  \includegraphics[width=0.97\textwidth]{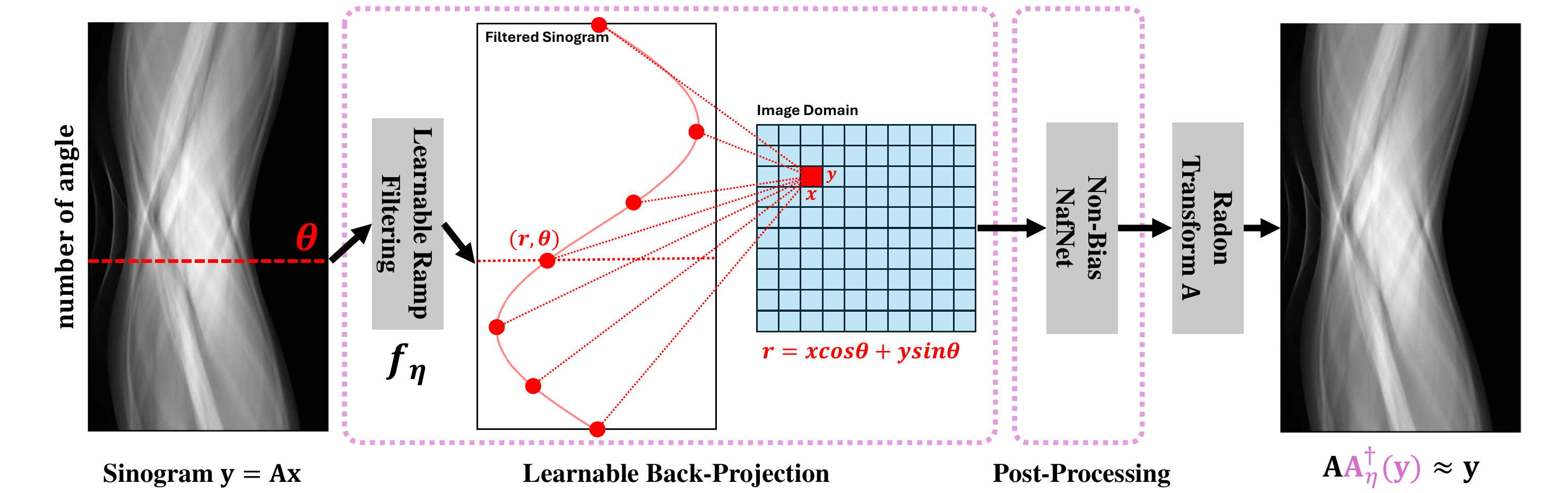}
  \hspace{-7pt}\vspace{-9pt}\caption{$\mathbf{A}^{\dagger}_{\eta}(\cdot)$ mainly consists of two blocks: The first is the learnable back-projection block, where we apply a learnable transformation to the Ramp filter and filter the input sinogram $\mathbf{Ax}$. Then, we back-project the image from the sinogram domain to the image domain. The second block uses a non-bias NafNet to perform post-processing on the back-projection results and outputs an estimation of the \textcolor{customPurple}{range-space content} \textcolor{customPurple}{$\mathbf{A}_\eta^{\dagger}(\mathbf{y})\in\mathbb{R}^{M}$} that satisfies $\mathbf{A}{\mathbf{A}_\eta^{\dagger}(\mathbf{y})}\approx\mathbf{y}$.}
  \label{fig:learn_pseudo}
\end{figure*}
As shown in Fig.~\ref{fig:learn_pseudo}, we implement this network using a Learnable Back-Projection block\footnote{Refer to Appendix~\ref{Appendix: Pseudo_inverse} for more details.} combined with a Non-bias NafNet~\cite{chen2022simple} for post-processing. The learnable back-projection block applies a learned Ramp filter $f_\eta$ to filter the input sinograms and subsequently back-projects them to the image domain, allowing us to resolve the dimensionality discrepancy between sinograms and images. The post-processing step refines the back-projection results, yielding more accurate estimations. To train such a model, we set the training objective as:
\begin{equation}\label{psu obj}
\begin{aligned}
& \ell_1(\eta )=\mathbb{E}_{\mathbf{x} \sim p(\mathbf{x}_0)}[\|\mathbf{y}-\mathbf{A} \hat{\mathbf{A}}_\eta^{\dagger} (\mathbf{y})\|],\\
& \ell_2(\eta) =\mathbb{E}_{\mathbf{y} \sim p(\mathbf{y})}[\|\hat{\mathbf{A}}_\eta^{\dagger}(\mathbf{y})- \hat{\mathbf{A}}_\eta^{\dagger} (\mathbf{A}\hat{\mathbf{A}}_\eta^{\dagger} (\mathbf{y}))\|],\\
& \ell_\text{pseudo}= (1-\omega) \ell_1(\eta) + \omega \ell_2(\eta),
\end{aligned}
\end{equation}
where $\omega\in [0,1]$ is a weighting factor controlling the trade-off between those two terms. This model can be optimized via stochastic gradient descent. Considering the consistency rectification is mainly enforced by $\ell_1(\eta )$, we recommend setting $\omega$ to 0.2 in practice to allocate more importance to $\ell_1(\eta )$. Alternatively, using only $\ell_1(\eta )$ during the early stages of training and fine-tuning the model with a balanced loss can achieve optimal training efficiency. \\

\noindent{\textbf{Error Compensation Through Range-Space Scaling:}} During the experiments, we noticed that the above iterative refinement sometimes tends to over-smooth the images, resulting in inferior reconstructions. We hypothesize that this may stem from the inaccurate estimation of the range-space content, which introduces an additional error $\mathbf{n}\in \mathbb{R}^{M}$ to the rectified clean image ${\mathbf{x}}_{0|t}^{\prime}$. Specifically, $\mathbf{n}$ satisfies,
\begin{equation}
    \hat{\mathbf{A}}_\eta^{\dagger} (\Delta\hat{\mathbf{y}}_{0|t})+\mathbf{n} = {\mathbf{A}}^{\dagger}\Delta\hat{\mathbf{y}}_{0|t},
\end{equation}
where $\Delta\hat{\mathbf{y}}_{0|t}=\mathbf{A}\hat{\mathbf{x}}_{0|t} -\mathbf{y}$. The problem is that when additional error, other than the anticipated noise $\sigma_t \mathrm{d} \widehat{\mathbf{w}}$, is introduced, the pretrained model will become unable to predict the intermediate clean images $\mathbf{x}_{0|t-1}$ accurately. Furthermore,  this error will accumulate iteratively, undermining the subsequent reverse diffusion process. To mitigate its impact, we rescale the range-space correction $\hat{\mathbf{A}}_\eta^{\dagger} (\Delta\hat{\mathbf{y}}_{0|t})$ by a rescaling factor $\gamma_t$, that is,
\begin{equation}
\mathbf{x}_{0|t}^{\prime}=\hat{\mathbf{x}}_{0|t}-\gamma_t \hat{\mathbf{A}}_\eta^{\dagger} (\Delta\hat{\mathbf{y}}_{0|t}).
\end{equation}
Here, we assign a hyperparameter $\alpha$ to control the rescaling operation and formulate $\gamma_t$ to be $\alpha\sigma_t\sqrt{\mathrm{d}t}/h_t$. This minor adjustment allows us to modify Eq.~\ref{xt-1} into: 
\begin{equation}\label{anti-n-sampling}
   \mathbf{x}_{t-1}=h_{t}\hat{\mathbf{x}}_{0| t} + \mathbf{c} + \sigma_t\sqrt{\mathrm{d}t}(\alpha\mathbf{n}+\bepsilon_t),\quad \bepsilon_t \sim \mathcal{N}(0, \mathbf{I}),
\end{equation}
where $\sqrt{\mathrm{d}t}\,\bepsilon_t = \mathrm{d} \widehat{\mathbf{w}}$ and $\mathbf{c}\in \mathbb{R}^M$ represents the constant terms that do not involve $\mathbf{x}_0$ and the noise. During sampling, this factor is crucial for stabilizing the reverse process. Choosing a smaller $\alpha$ can minimize the impact of $\mathbf{n}$, but this at the same time also weakens the range-space correction, resulting in worse data consistency. 

Apart from the rescaling, we empirically found that inserting unguided iterations,  i.e., those without the data consistency term, between guided iterations can effectively prevent the accumulation of errors. In practice, we skip the consistency term at specific iterations, with the frequency of skipping regulated by an extra hyperparameter $\beta$. In this paper, we focus on the LACT problem, where $\mathbf{A}^\dagger$ needs to be estimated with a neural network. The skipping trick mentioned above is not required for other simpler IR tasks, where the pseudo-inverse can be numerically calculated, such as image inpainting and colorization. The algorithm details are shown in Algo.~\ref{RN-SDE sampling algo}.\\
\begin{algorithm}[htb]
\caption{Sampling Process of RN-SDE}\label{RN-SDE sampling algo}
    \begin{algorithmic}[1]
    \Require The low-fidelity reconstruction $\bmu$, the degradation operator $\mathbf{A}$, the learned pseudo-inverse ${\mathbf{A}}_\eta^{\dagger}(\cdot)$, the rescaling factor $\alpha$, and the skipping factor $\beta$
    \State $\mathbf{x}_\T \sim \mathcal{N}(\bmu,\lambda^2)$
    \For{$t = \T,\ldots,1$}
        \State $  \hat{\mathbf{x}}_{0|t} = - \frac{g_t}{h_t}\mathbf{x}_t + \frac{\sigma_t^2}{h_t} \hat{\mathbf{s}}_\phi(\mathbf{x}_t, \bmu, t)  \mathrm{d} t + (1+\frac{g_t}{h_t})\bmu$, \vspace{+3pt}
            \If{$t \mod \beta \neq 0$}
            \State ${\mathbf{x}}_{0|t}^{\prime}=\hat{\mathbf{x}}_{0|t}-\gamma_t \hat{\mathbf{A}}_\eta^{\dagger} (\Delta\hat{\mathbf{y}}_{0|t})$, \quad  $\gamma_t=\frac{\alpha\sigma_t\sqrt{\mathrm{d}t}}{h_t}$
            \Else  
            \State ${\mathbf{x}}_{0|t}^{\prime}=\hat{\mathbf{x}}_{0|t}$, \vspace{+1pt}
            \EndIf
        \State ${\mathbf{x}}_{t-1} \sim p(\mathbf{x}_{t-1} | \mathbf{x}_t, {\mathbf{x}}_{0|t}^{\prime})$, \quad  \quad \quad \quad $\Rightarrow$ Equation~\ref{anti-n-sampling}
    \EndFor
    \State \Return $\mathbf{x}_0$
    \end{algorithmic}
\end{algorithm}

\noindent\textbf{Time-Travel Trick:} Except for the above iterative refinement pipeline,~\cite{wang2023zero} proposes a time-travel trick to address the \textit{inferior Realness} caused when the range-space content are too local or uneven. As the name suggests, during sampling ($\T\rightarrow 0$), we travel backward from a chosen time-step, $t$, to the past, $t + l$. Due to the iterative refinement process, the \textbf{past} state $\mathbf{x}_{t+l}$, which we sampled from $p(\mathbf{x}_{t+l}| \mathbf{x}_t)$, will be more accurate than the real past state. Therefore, as we return to the current time-step $t$ again through the RN-SDE sampling iteration (Eqs.~\ref{clean inter-states},~\ref{rectification},~\ref{xt-1}), we obtain an improved estimate of $\mathbf{x}_t$. 

This process is controlled by three hyperparameters: The length of time we travel back each time, $l$, the number of times we repeat the same time interval, $r$, and the total number of diffusion steps, $\T$. Typically, the actual number of time steps that the sampling with time traveling goes through, $\T_{tt}$, can be measured by the equation below:
\begin{equation}\label{eq: tt iter}
\T_{tt} = \T + 2 l (r - 1) \left\lfloor \frac{\T - 1}{l} \right\rfloor + 1.
\end{equation}
When $r=1$ and $l=1$, the sampling process with time traveling will be equivalent to the normal sampling process. However, a major drawback of this trick is that even when we set the minimum repetition time, i.e., $r=2$, the actual sampling step number $\T_{tt}$ will still be almost three times the original $\T$. Fortunately, using MR-SDE, we can keep $\T_{tt}$ within a reasonable range.

\section{Experimental Results}\label{sec: experiments}
In this section, we present our experimental setup and results, comprehensively evaluating our proposed method against several SOTA methods. In quantitatively comparing different methods, we consider several widely used metrics, including the standard distortion metrics peak signal-to-noise ratio (PSNR) and structural similarity index measure (SSIM), as well as the popular perceptual metric, Learned Perceptual Image Patch Similarity (LPIPS)~\cite{zhang2018unreasonable}. Considering the computational and time costs, we primarily conducted experiments using the ChromSTEM dataset, while the public C4KC-KiTS~\cite{heller2019kits19} dataset was also used as a complement in Sec~\ref{experiment: compare} to further show the effectiveness and generalizability of our method.

\subsection{Dataset And Preprocessing Procedures}
\noindent\textbf{ChromSTEM Dataset:} The \textit{Chromatin Scanning Transcription Electron Microscopy (ChromSTEM)} dataset is synthesized using the SR-EV~\cite{carignano2024local} model, designed by the Northwestern University Center for Chromatin NanoImaging in Cancer. Based on current knowledge of chromatin's structural and statistical characteristics, SR-EV effectively simulates the three-dimensional structure of chromatin, demonstrating a high degree of visual similarity to real data samples. The dataset used here comprises $1196$ synthesized 3D chromatin structures of stem cells. Each chromatin structure consists of $256$ slices that are $50\times256$ pixels in size, with 2 nm spatial resolution. To facilitate network training, a random sample of approximately $17,200$ sub-volumes with 16 channels from these 3D structures are selected and zero-padded to $64\times 256$. Their corresponding sinograms are limited to the size of $91\times 256$ to fully simulate the real ChromSTEM imaging scenario, where a projection is obtained every 2\degree across the [0\degree,180\degree] interval. Finally, a random selection of 24 sub-volumes from distinct chromatin structures is chosen as the testing set, while the remaining sub-volumes are utilized for training. Access to the ChromSTEM dataset is available in our project repository.\\
\begin{figure*}[htb]
  \centering
  \includegraphics[width=\textwidth]{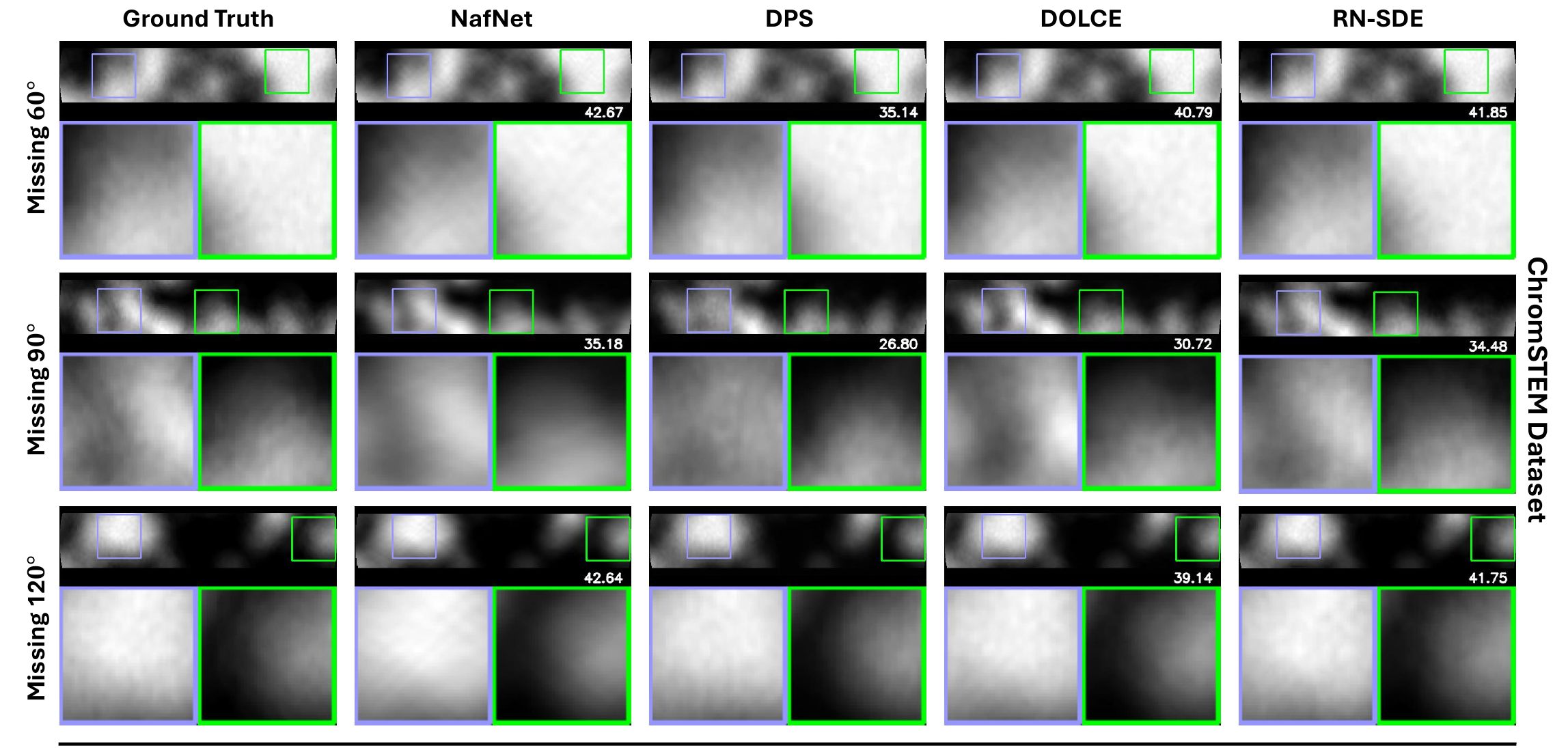}
  \hspace{-7pt}\vspace{-5pt}\caption{Visual evaluation of limited angle tomographic reconstruction on the ChromSTEM test set using FBP, NafNet, DPS, DOLCE, and our RN-SDEs. From top to bottom, we present the visualization of a typical reconstruction result for $\theta_{\rm miss} \in$ \{60\degree, 90\degree, 120\degree\}. The bottom right corner of each example shows the corresponding PSNR metric compared to the Ground Truth (\textbf{first column}).}
  \label{fig:chromstem}
\end{figure*}
\noindent\textbf{C4KC-KiTS Dataset:} To better compare our method with other SOTA methods, such as DOLCE~\cite{liu2023dolce}, we utilized the Kidney CT scans from the publicly available \textit{2019 Kidney and Kidney Tumor Segmentation Challenge (C4KC-KiTS)}~\cite{heller2019kits19} dataset. The C4KC-KiTS dataset contains approximately $70,000$ 2D body scans of size $512\times512$, covering a range of anatomical regions from the chest to the pelvis. Unfortunately, since DOLCE does not publish the complete partitioning of the training and test sets, we cannot compare their results directly. To make a fair comparison, we adhered to the DOLCE projection setup, resulting in sinograms of size $720 \times 512$, i.e., a uniform scanning at every 0.25\degree. Besides, in this paper, the CT intensity is clipped to the range of $[-250, 500]$ HU (Hounsfield Unit) following~\cite{heller2021state}, and then normalized to $[0, 1]$. We randomly selected 24 body scans from different patients for testing and used the remaining images for training.
\begin{figure*}[htb]
  \centering
  \includegraphics[width=\textwidth]{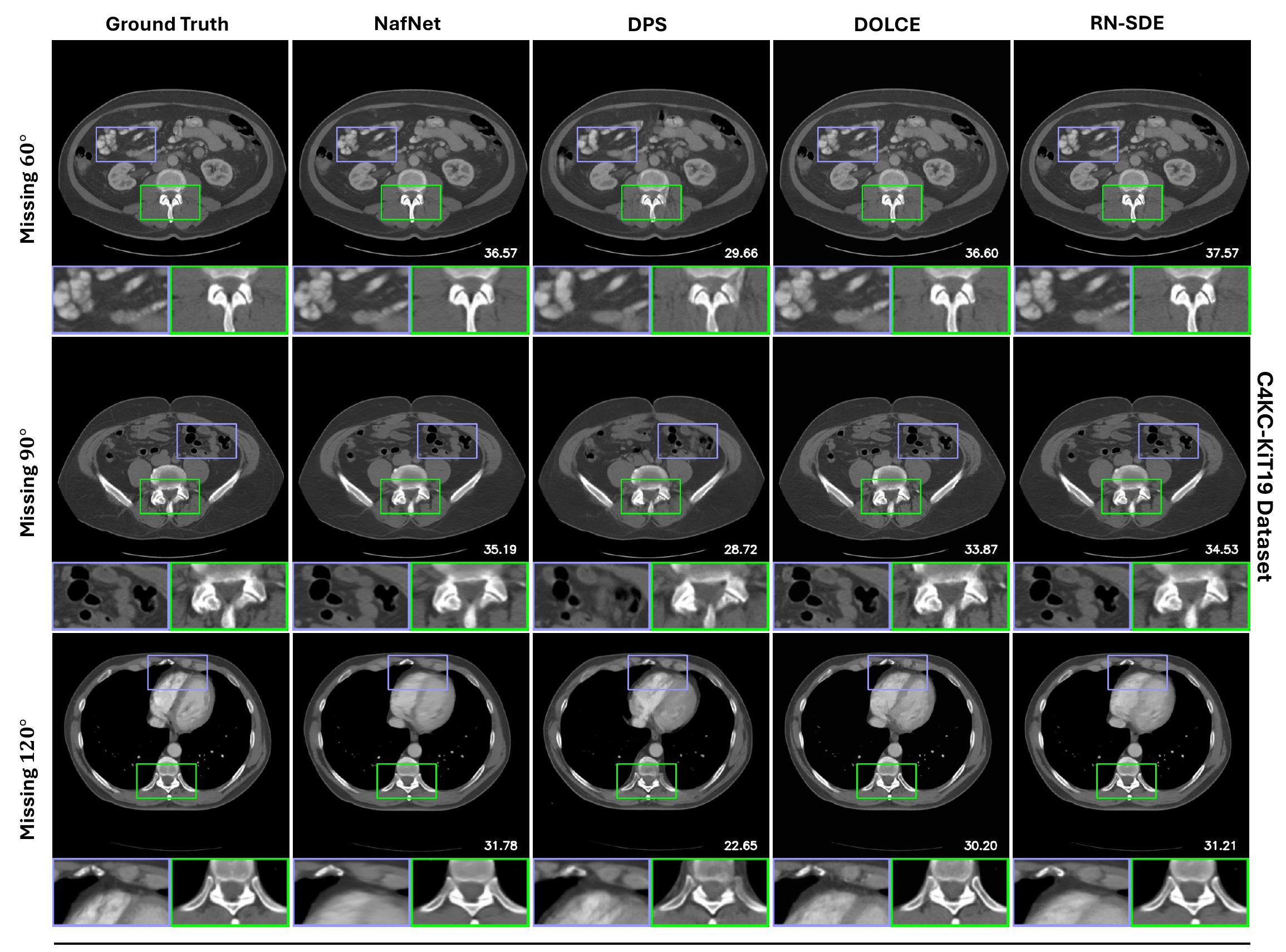}
  \hspace{-7pt}\vspace{-5pt}\caption{Visual evaluation of limited angle tomographic reconstruction on the C4KC-KiTS dataset using FBP, NafNet, DPS, DOLCE, and our RN-SDE. From top to bottom, we present the visualization of a typical reconstruction for $\theta_{\rm miss} \in$ \{60\degree, 90\degree, 120\degree\} respectively. The bottom right corner of each example shows the corresponding PSNR metric compared to the Ground Truth (\textbf{first column}).}
  \label{fig:c4kc}
\end{figure*}

\subsection{Implementation Details}
The model backbone utilized in RN-SDE is the Cond-NafNet~\cite{luo2023refusion}. For the two datasets mentioned above, we evaluate our proposed RN-SDE based on three missing angles, i.e., $\theta_{\rm miss} \in$ \{60\degree, 90\degree, 120\degree\}, training individual models for each LACT scenario. For the noise schedules of diffusion, we follow the default setting of~\cite{luo2023image}, progressively adding noise at each step according to a cosine schedule until reaching the terminal state, $\mathbf{x}_\T$. Specifically, for the ChromSTEM dataset, to validate the impact of using different $\bmu$ on the final reconstruction results, we experimented with different time steps $\T$ (as shown in Table~\ref{tab:differentT}) and selected the one that performed best on the LPIPS metric for subsequent experiments. For the C4KC-KiTS dataset, we used the same $\T$ to demonstrate that the choice of $\T$ does not significantly impact the final reconstruction. For the inference process, given the advantages of RN-SDE, we need far fewer time steps to achieve competitive results than the normal DDPM diffusion. Therefore, we adopted the most widely used DDPM~\cite{ho2020denoising} sampling method, i.e., using the same number of diffusion steps $\T$ as in training, to obtain the best reconstruction results. For training, all models are implemented based on PyTorch and trained on a single Nvidia Quadro RTX 8000 GPU and an Intel Xeon Gold 6226R CPU with single precision ($\text{float 32}$). We use the AdamW optimizer while the learning rate is initialized to be $5\times 10^{-4}$ and controlled by the Cosine Annealing scheduler. The batch size is set to $8$ for all experiments. 

\subsection{Ablation Study}
\noindent\textbf{Ablation Study on Time Steps with Mean-Reverting.} We highlighted that the mean-reverting diffusion strategy effectively reduces the number of steps required during the actual reverse diffusion process. To validate the effectiveness of this strategy, we retrained the network using different values of $\T$ across three distinct LACT scenarios on the ChromSTEM dataset, conducting sampling without applying the RNSD-based rectification. We report the average result of the 10 sampling runs in Table~\ref{tab:differentT}. As anticipated, we observe a significant improvement in distortion-based metrics as $\T$ increases. However, this improvement noticeably slows once $\T$ approaches 200. From a perceptual perspective, the best performance typically occurs at $\T = 100$, indicating that this value is sufficient to ensure the convergence of the diffusion process. Considering these factors, we set $\T$ to $200$ by default in all subsequent experiments unless otherwise specified.

\begin{table}[htbp]
\caption{Average PSNR/SSIM/LPIPS Results of RN-SDE (w/o RNSD Rectification) Trained with Different Time Steps on ChromSTEM Test images}\label{tab:differentT} 
\centering
\resizebox{0.5\textwidth}{!}{
\setlength{\tabcolsep}{3.5pt}
\begin{tabular}{lccccccccc}
\toprule
\multirow{2}{*}{\textbf{T}} & \multicolumn{3}{c}{\textbf{60\degree}}   & \multicolumn{3}{c}{\textbf{90\degree}}  & \multicolumn{3}{c}{\textbf{120\degree}} \\ 
\cmidrule(lr){2-4} \cmidrule(lr){5-7} \cmidrule(lr){8-10} 
& \textbf{\metric PSNR$\uparrow$}  & \textbf{\metric SSIM$\uparrow$}   & \textbf{\metric LPIPS$\downarrow$}  & \textbf{\metric PSNR$\uparrow$}  & \textbf{\metric SSIM$\uparrow$}   & \textbf{\metric LPIPS$\downarrow$}  & \textbf{\metric PSNR$\uparrow$}  & \textbf{\metric SSIM$\uparrow$}   & \textbf{\metric LPIPS$\downarrow$} \\ \midrule
\textbf{50}&    36.50 & 0.905 & 0.0932 &   36.01 & 0.889 & 0.1045&   33.27& 0.860& 0.1138\rule{0pt}{1.9ex}\\
\textbf{100}&   41.77 & 0.972 & 0.0097 &   38.96 & 0.955 & 0.0072&   35.07& 0.931& 0.0112\rule{0pt}{1.9ex}\\
\textbf{200}&   42.08 & 0.975 & 0.0062 &   39.12 & 0.958 & 0.0080&   35.17& 0.933& 0.0125\rule{0pt}{1.9ex} \\
\textbf{500}&   42.24 & 0.976 & 0.0110 &   39.20 & 0.959 & 0.0109&   35.19& 0.935& 0.0130\rule{0pt}{1.9ex}\\
\textbf{1000}&  42.45 & 0.976 & 0.0162 &   39.16 & 0.957 & 0.0229&  35.20 & 0.935 & 0.0163\rule{0pt}{1.9ex}\\ \bottomrule
\end{tabular}
}
\end{table}

\noindent\textbf{Ablation Study with different $\bmu$ and Incorporating RNSD Data-Consistency.} As previously mentioned, in RN-SDE, we utilize the accurate low-frequency information provided by $\mathbb{E}[\mathbf{x}|\mathbf{y}]$ to control and guide the mean-reverting diffusion process, which undoubtedly improves the quality of image restoration. However, this requires additional training of a NafNet, which incurs extra time costs. We question whether a more accessible modality could be used to define $\bmu$ while entrusting the data consistency to RNSD rectification. For this purpose, we further evaluate our method by adopting FBP reconstruction as $\bmu$ for mean-reverting. For different $\bmu$, we experimented with two situations: with (w/) and without (w/o) the RNSD-based data-consistency measurement  ($\mathbf{A}^{\dagger}$) provided in Eq.~\ref{rectification}. For simplicity, we refer to the group of experiments using FBP reconstruction as $\bmu$ to be \textbf{FBP-MR}, and those using NafNet reconstruction as \textbf{NAF-MR}.

\begin{table}[htb]
\caption{Ablation Study on Mean-Reverting Strategies and RNSD-Based Rectification}\label{tab: fbp mr-sde}
\centering
\resizebox{0.49\textwidth}{!}{
\setlength{\tabcolsep}{2.5pt}
\vspace{-12pt}\begin{tabular}{lccccccccc}
\toprule
\textbf{Angle} &
  \multicolumn{3}{c}{\textbf{60\degree}} &
  \multicolumn{3}{c}{\textbf{90\degree}} &
  \multicolumn{3}{c}{\textbf{120\degree}} \\ \cmidrule(lr){2-4} \cmidrule(lr){5-7} \cmidrule(lr){8-10}
\textbf{Metric} &
  \metric{PSNR$\uparrow$} &
  \metric{SSIM$\uparrow$} &
  \metric{LPIPS$\downarrow$} &
  \metric{PSNR$\uparrow$} &
  \metric{SSIM$\uparrow$} &
  \metric{LPIPS$\downarrow$} &
  \metric{PSNR$\uparrow$} &
  \metric{SSIM$\uparrow$} &
  \metric{LPIPS$\downarrow$} \\ 
\midrule
\multicolumn{9}{l}{\emph{\textbf{NAF-MR}: NafNet Reconstruction As $\bmu$}} \rule{0pt}{1.5ex}\\
\midrule
\textbf{w/o $\mathbf{A}^{\dagger}$}

& 42.51 & 0.977 & 0.0078
& 39.52 & 0.961 & 0.0097
& 35.46 & 0.939 & 0.0144 \\
\textbf{w/ $\mathbf{A}^{\dagger}$}
& 43.94 & 0.982 & 0.0069
& 40.43 & 0.965 & 0.0086
& 35.62 & 0.939 & 0.0143\\
\midrule
\multicolumn{9}{l}{\emph{\textbf{FBP-MR}: FBP Reconstruction As $\bmu$}} \rule{0pt}{1.5ex}\\
\midrule
\textbf{w/o $\mathbf{A}^{\dagger}$} 
& 35.13 & 0.936 & 0.0111
& 28.46 & 0.880 & 0.0269
& 21.73 & 0.756 & 0.0726 \\
\textbf{w/ $\mathbf{A}^{\dagger}$}  
& 42.73 & 0.975 & 0.0053
& 38.73 & 0.955 & 0.0033
& 29.44 & 0.872 & 0.0328 \\ \bottomrule
\end{tabular}
}
\end{table}

\begin{figure}[htb]
  \centering
  \hspace{-8pt}\includegraphics[width=0.49\textwidth]{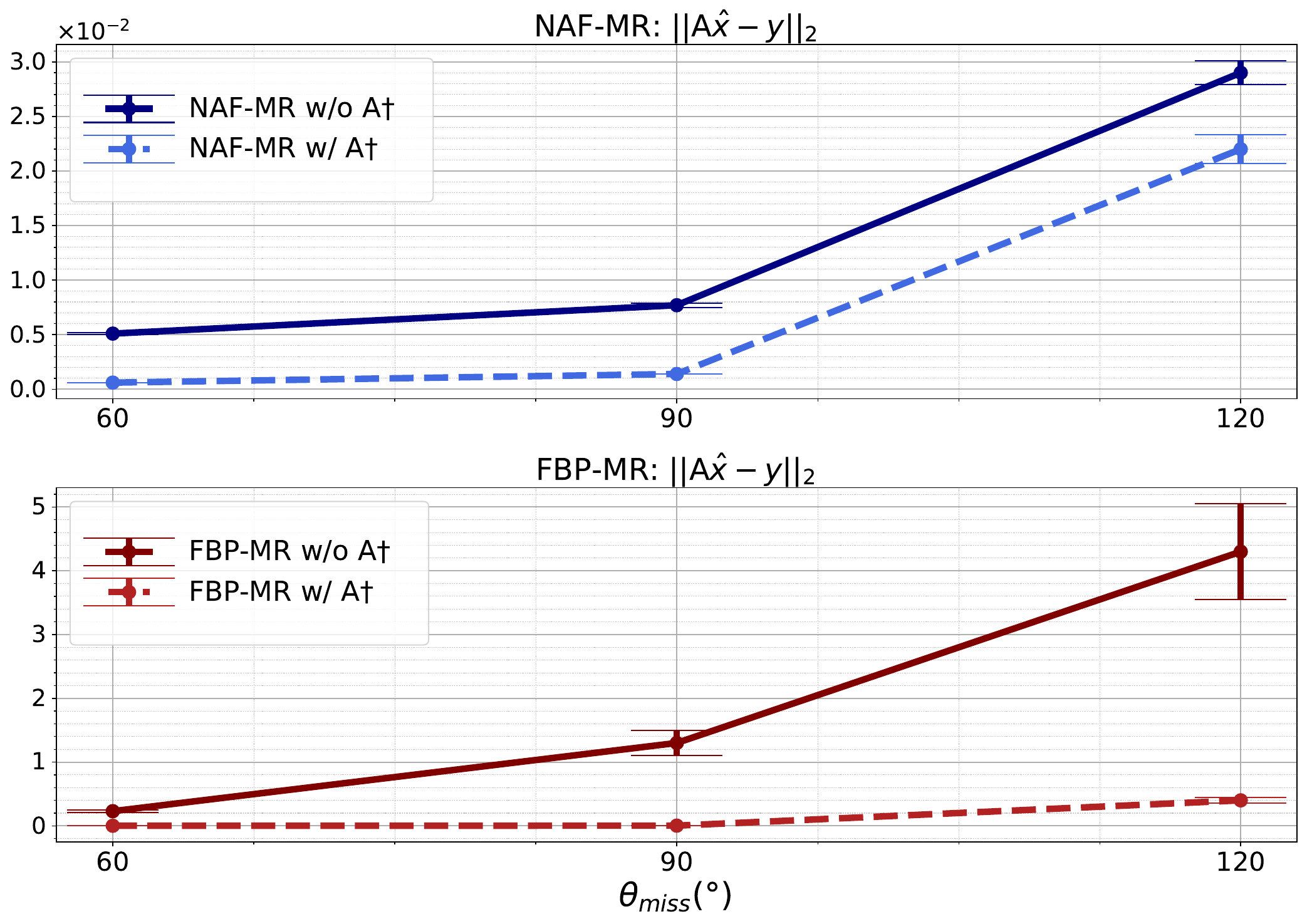}
 \caption{
        Visualization of the data consistency error: $\|\mathbf{A} \mathbf{\hat{x}}-\mathbf{y}\|_2$. Compared to \textcolor[rgb]{0.698, 0.133, 0.133}{FBP-MR}, \textcolor[rgb]{0.25, 0.412, 0.882}{NAF-MR} shows a smaller consistency error, but this gap narrows after the introduction of RNSD rectification. Notably, including RNSD-based rectification significantly enhances and stabilizes reconstruction performance, as evidenced by a considerable reduction in the mean and standard deviation.
    }
  \label{fig: vis table 2}
\end{figure}

As shown in Table~\ref{tab: fbp mr-sde}, without $\mathbf{A}^{\dagger}$, we observed a remarkable decline in reconstruction quality for FBP-MR compared to the NAF-MR. This trend is evident across all metrics and is particularly pronounced in severe LACT scenarios. We attribute this to the sensitivity of MR-SDE to the choice of $\bmu$. After applying $\mathbf{A}^{\dagger}$, although the results for PSNR and SSIM in the FBP-MR group remain inferior to those of the NAF-MR group, the perceptual quality, as measured by LPIPS, has surprisingly surpassed that of the NAF-MR group. Meanwhile, it is worth mentioning that \textbf{FBP-MR outperforms the current SOTA method, DOLCE},  on the $60\degree$ and $90\degree$ LACT scenarios (Table~\ref{tab: example}). Another noticeable change comes from the data consistency term, $\|\mathbf{A} \mathbf{\hat{x}}-\mathbf{y}\|_2$, for which we plot its mean and standard deviation in Fig.~\ref{fig: vis table 2}. After applying RNSD rectification, the mean and variance notably decrease, indicating that the possible solution $\mathbf{\hat{x}}$ has been constrained to a smaller range, resulting in more stable and accurate reconstructions.

Those observations highlight that the mean-reverting strategy is highly sensitive to the choice of $\bmu$ and underscore the effectiveness and necessity of incorporating $\mathbf{A}^{\dagger}$ for addressing various LACT scenarios. For the subsequent experiments, given the final application scenario of this study, we will still use NAF-MR as the default scheme.

\subsection{Evaluation and Comparison on Two Datasets}\label{experiment: compare}
\begin{table*}[htbp]
\caption{Average PSNR/SSIM/LPIPS Results for Several Methods on ChromSTEM and C4KC-KiTS Test Images}\label{tab: example}
\centering
\resizebox{\textwidth}{!}{
\setlength{\tabcolsep}{2.5pt}
\begin{tabular}{lcccccccccccccccccc}
\toprule
\textbf{Dataset} \rule{0pt}{2.2ex} & \multicolumn{9}{c}{\textbf{ChromSTEM}} & \multicolumn{9}{c}{\textbf{C4KC-KiTS}} \\ 
\cmidrule(lr){2-10} \cmidrule(lr){11-19}
\textbf{Angle} \rule{0pt}{1.8ex} &
  \multicolumn{3}{c}{\textbf{60}\degree} & \multicolumn{3}{c}{\textbf{90}\degree} & \multicolumn{3}{c}{\textbf{120}\degree} &
  \multicolumn{3}{c}{\textbf{60}\degree} & \multicolumn{3}{c}{\textbf{90}\degree} & \multicolumn{3}{c}{\textbf{120}\degree} \\
\cmidrule(lr){2-4} \cmidrule(lr){5-7} \cmidrule(lr){8-10} \cmidrule(lr){11-13} \cmidrule(lr){14-16} \cmidrule(lr){17-19}
\textbf{Metric} &
  \textbf{\metric PSNR$\uparrow$} & \textbf{\metric SSIM$\uparrow$} & \textbf{\metric LPIPS$\downarrow$} &
  \textbf{\metric PSNR$\uparrow$} & \textbf{\metric SSIM$\uparrow$} & \textbf{\metric LPIPS$\downarrow$} &
  \textbf{\metric PSNR$\uparrow$} & \textbf{\metric SSIM$\uparrow$} & \textbf{\metric LPIPS$\downarrow$} &
  \textbf{\metric PSNR$\uparrow$} & \textbf{\metric SSIM$\uparrow$} & \textbf{\metric LPIPS$\downarrow$} &
  \textbf{\metric PSNR$\uparrow$} & \textbf{\metric SSIM$\uparrow$} & \textbf{\metric LPIPS$\downarrow$} &
  \textbf{\metric PSNR$\uparrow$} & \textbf{\metric SSIM$\uparrow$} & \textbf{\metric LPIPS$\downarrow$} \\
\midrule
\textbf{FBP}
& 18.30 & 0.573 & 0.2452
& 14.19 & 0.463 & 0.3038
& 8.26 & 0.275 & 0.4519
& 16.22 & 0.316 & 0.1914
& 13.92 & 0.255 & 0.2771
& 11.34 & 0.178 & 0.3688 \\
\textbf{TV}~\cite{chambolle2011first} 
& 27.64 & 0.841 & 0.1101
& 23.32 & 0.758 & 0.1536
& 20.52 & 0.692 & 0.1967
& 23.37 & 0.489 & 0.2959
& 21.37 & 0.379 & 0.3380
& 19.48 & 0.330 & 0.3775 \\\midrule
\textbf{NafNet}~\cite{chen2022simple}         
& \hlblue{44.31} & \hlblue{0.985} & 0.0320
& \hlblue{41.09} & \hlblue{0.975} & 0.0525
& \hlblue{36.35} & \hlgreen{0.957} & 0.0684
& \hlgreen{37.57} & \hlgreen{0.971} & 0.0518
& \hlgreen{33.68} & \hlgreen{0.949} & 0.0776
& \hlgreen{30.29} & \hlgreen{0.918} & 0.1037 \\ \midrule
\textbf{DPS}~\cite{chung2023diffusion}         
& 38.76 & 0.951 & \hlpurple{0.0123}
& 30.10 & 0.850 & 0.0414
& 26.48 & 0.849 & \hlpurple{0.0370}
& 29.78 & 0.858 & 0.0642
& 25.39 & 0.820 & 0.0954
& 21.60 & 0.764 & 0.1415 \\
\textbf{DDNM+}~\cite{wang2023zero}             
& 35.39 & 0.907 & 0.0145
& 31.69 & 0.859 & \hlpurple{0.0348}
& 27.18 & 0.813 & 0.0449
& 21.31 & 0.655 & 0.2087
& 20.26 & 0.637 & 0.2344
& 18.29 & 0.601 & 0.2939 \\
\textbf{DOLCE}~\cite{liu2023dolce}           
& 42.61 & 0.978 & \hlblue{0.0119}
& 38.47 & 0.959 & \hlblue{0.0140}
& 31.36 & 0.915 & \hlblue{0.0280}
& 35.80 & 0.956 & \hlgreen{0.0180}
& 32.41 & 0.928 & \hlgreen{0.0325}
& 28.57 & 0.879 & \hlgreen{0.0479} \\ \midrule
\textbf{RN-SDE} \rule{0pt}{1.8ex}     
& \hlpurple{43.94} & \hlpurple{0.982} & \hlgreen{0.0069}
& \hlpurple{40.43} & \hlpurple{0.965} & \hlgreen{0.0086}
& \hlpurple{35.62} & \hlpurple{0.939} & \hlgreen{0.0143}
& \hlpurple{36.86} & \hlpurple{0.960} & \hlblue{0.0257}
& \hlpurple{32.89} & \hlpurple{0.936} & \hlblue{0.0393}
& \hlpurple{29.53} & \hlpurple{0.902} & \hlblue{0.0629} \\
\textbf{RN-SDE SA}        
&  \hlgreen{45.36} & \hlgreen{0.987} & 0.0238
&  \hlgreen{41.92} & \hlgreen{0.976} & 0.0390
&  \hlgreen{36.55} & \hlblue{0.954} & 0.0539
& \hlblue{37.38} & \hlblue{0.966} & \hlpurple{0.0464}
& \hlblue{33.61} & \hlblue{0.948} & \hlpurple{0.0713}
& \hlblue{30.00} & \hlblue{0.915} & \hlpurple{0.0980} \\ \bottomrule
\end{tabular}
}
\\
\raggedright 
{\vspace{2pt}\hspace{3pt} 
\scriptsize Note: The \hlgreen{best}, \hlblue{second-best} and the \hlpurple{third-best} values for each metric have been highlighted with colors.} 
\end{table*}

In this section, we compare RN-SDE with six selected LACT reconstruction methods. Table~\ref{tab: example} presents the ten-run average PSNR, SSIM, and LPIPS results for the selected methods, using test images from both the ChromSTEM and C4KC-KiTS datasets. The chosen methods are categorized into three major groups: classical methods (e.g., FBP, TV~\cite{chambolle2011first}), MMSE-based IR methods (e.g., NafNet~\cite{chen2022simple}), and diffusion-based IR methods (DPS~\cite{chung2023diffusion}, DDNM+~\cite{wang2023zero}, and DOLCE~\cite{liu2023dolce}). 

The TV reconstructions were implemented using TomoPy~\cite{gursoy2014tomopy}, a publicly available CT reconstruction toolkit. NafNet refers to the original framework without embedding input, and its input is also the FBP reconstruction. As NafNet has been shown to outperform the standard UNet in various image restoration tasks, we have opted not to include UNet in our comparisons. DDNM+ and DPS are two image restoration algorithms based on unconditioned diffusion. They were originally proposed to address general inverse problems, but not specifically designed for the LACT problem. Notably, DDNM+ also utilizes the RNSD approach to guide the reverse diffusion, hence we used the same pre-trained ${\mathbf{A}}_\eta^{\dagger}(\cdot)$ as for our RN-SDE. DOLCE is claimed to be the current SOTA method for the LACT problem. Unlike the previous two methods, DOLCE leverages low-quality reconstruction as the condition during training to endorse better reconstruction quality. It is worth noting that all three diffusion-based methods use DDPM~\cite{ho2020denoising} as the foundational diffusion framework, while our RN-SDE is essentially based on stochastic differential equations. Furthermore, for our RN-SDE model, we followed the sampling average (SA) outlined in \cite{whang2022deblurring}, averaging multiple samples from RN-SDE to approximate $\mathbb{E}[\mathbf{x}|\mathbf{y}]$. This enhances our reconstruction in terms of the distortion metrics, with the tradeoff of perceptual quality. We denote this as \textbf{RN-SDE SA} in Table~\ref{tab: example}. The main purpose is to ensure a fair comparison between diffusion-based methods and MMSE-based methods like NafNet. For the ChromSTEM dataset, we conduct a grid search over the data consistency parameters for DPS, DDNM+, and DOLCE and adopt the same DDPM diffusion setting. For the C4KC-KiTS dataset, we directly copy the diffusion setting provided in DOLCE to train both conditional and unconditional DDPM diffusion models. We also perform a grid search over the data consistency parameters for DPS and DDNM+. Importantly, we train different models for different LACT scenarios and always use the standard DDPM sampling scheme without skipping. 

As expected, in terms of perceptual quality, as indicated by the LPIPS metric, all diffusion-based reconstruction methods achieve satisfactory results. Among them, RN-SDE and DOLCE stand out, significantly outperforming others, with each showing competitive performance across the two datasets, and the differences between them are negligible. Furthermore, in terms of PSNR and SSIM, RN-SDE outperforms DOLCE across all experimental cases on both datasets. For NafNet and RN-SDE, as previously mentioned, a direct comparison between these two experiments is not entirely fair. Nevertheless, the difference between RN-SDE and NafNet remains minimal. In fact, concerning PSNR and SSIM, the difference between RN-SDE and NafNet across all experimental settings stays within 1 dB for PSNR and 0.02 for SSIM. However, in terms of perceptual quality (LPIPS), RN-SDE demonstrates a significant improvement over NafNet. When comparing RN-SDE SA to NafNet, RN-SDE SA achieves a clear advantage on the ChromSTEM dataset, while for the C4KC-KiTS dataset, this gap narrows to 0.3 dB for PSNR and 0.005 for SSIM.
\\

\noindent\textbf{Model Complexity and Runtime Efficiency.} In Table~\ref{tab: time}, we compared the computational complexity of the sampling process in four diffusion-based models across four aspects: FLOPs, number of parameters, and time consumption, using both the ChromSTEM and C4KC-KiTS datasets. We reported the total time consumption for the sampling process, along with the corresponding number of iterations required. The iterations for RN-SDE include Time-Travel, which can be measured by Eq.~\ref{eq: tt iter}. In the ChromSTEM dataset, we set the travel length to $l=4$ and the travel repetition to $r=2$, while for the C4KC dataset, we adjusted the travel length to $l=8$. The batch size was set as $1$. In DOLCE, we used our own implementation of the Radon forward/backward transform, which may lead to differences in execution efficiency. Therefore, we must emphasize that the provided \textbf{times are approximate and intended for reference only.}

\begin{table}[ht]
\caption{Comparison of Model Complexity and Runtime Efficiency, Experiments are based on a single Nvidia Quadro RTX 8000 GPU, and an Intel(R) Xeon(R) Gold 6226R CPU}\label{tab: time}
\centering
\resizebox{0.49\textwidth}{!}{
\setlength{\tabcolsep}{2.5pt}
\vspace{-7pt}\begin{tabular}{llccrr}
\toprule
\textbf{Method} \rule{0pt}{2.0ex} & \textbf{Backbone} & \textbf{\#GMacs} & \textbf{\#Params} & \multicolumn{1}{c}{\textbf{Time}} & \multicolumn{1}{c}{\textbf{Iters}}\\ 
\midrule
\multicolumn{6}{l}{\textit{ChromSTEM Dataset}} \rule{0pt}{1.5ex}\\
\midrule
\textbf{DPS} & UNet & 4.94e+4 & 93.61 M & 171 s & 1000\\
\textbf{DDNM+} & UNet & 4.93e+4 & 93.61 M & 126 s & 1000\\
\textbf{DOLCE} & CondUNet & 5.34e+4 & 93.63 M & 4634 s & 1000\\
\textbf{RN-SDE} & CondNafNet & 6.02e+3 & 70.08 M & 66 s & 593\\
\midrule
\multicolumn{6}{l}{\textit{C4KC-KiTS Dataset}} \rule{0pt}{1.5ex}\\
\midrule
\textbf{DPS} & UNet & 7.20e+5 & 94.53 M & 951 s & 2000\\
\textbf{DDNM+} & UNet & 7.18e+5 & 94.53 M & 1011 s & 2000\\
\textbf{DOLCE} & CondUNet & 8.39e+5 & 94.53 M & 10875 s & 2000\\
\textbf{RN-SDE} & CondNafNet & 9.92e+4 & 85.72 M & 106 s & 585\\
\bottomrule
\end{tabular}
}
\\
\raggedright 
{\vspace{2pt}\hspace{4pt} 
\scriptsize \parbox{\dimexpr\linewidth-8pt}{Note: Different implementations may result in variations in code execution efficiency, the times presented are approximate and intended for reference only.}}
\end{table}
The table highlights that RN-SDE offers a clear advantage over DOLCE in terms of both computational complexity and runtime efficiency. Specifically, RN-SDE requires significantly fewer FLOPs and parameters, completing the sampling process in a fraction of the time compared to DOLCE, which exhibits the highest time consumption on both datasets. Notably, RN-SDE achieves these improvements while also outperforming DOLCE in key performance metrics such as PSNR, SSIM, and LPIPS, making it a more efficient and effective choice overall.
\\

\noindent\textbf{Visual Evaluation.} We compare several representative visual results of our proposed RN-SDE with those from NafNet, DPS, and DOLCE. As shown in Figs.~\ref{fig:c4kc} \&~\ref{fig:chromstem}, several notable characteristics can be observed: for both datasets, while the NafNet reconstructions perform well in terms of PSNR and SSIM metrics, the visual results show a clear lack of high-frequency details, leading to visible blurring, particularly in cases with $120^\circ$ angle missing. In contrast, the visual results of the other three diffusion models all exhibit satisfactory perceptual quality, which can be attributed to the nature of diffusion models that directly model the entire image distribution and sample from the posterior. Among them, DPS shows weaker data consistency compared to DOLCE and RN-SDE. We hypothesize that this disparity is partially due to DPS not using low-quality reconstructions as a condition to guide the diffusion process.
As for DOLCE and RN-SDE, their visual results show no significant differences. As mentioned earlier, the primary difference between these models lies in their complexity and the time consumption during the sampling process, which will not be elaborated further here.

\section{Discussion \& Conclusion}
\label{sec:conclusion_discussion}
In this paper, we address the problem of LACT reconstruction under noise-free conditions. Building upon diffusion stochastic differential equations, we introduce RN-SDEs, a novel NafNet-based diffusion framework that leverages pre-trained mean-reverting SDEs as image priors for high-quality CT reconstruction. Specifically, we use raw reconstructions as the \textit{mean} to initialize and condition the reverse diffusion process, which has been proven to remarkably reduce the diffusion time steps required for convergence and achieve better perceptual quality. To further enforce data consistency, we incorporate an additional rectification step based on RNSD during each iteration, which involves replacing the range-space content of the intermediate clean estimation while preserving the null-space content intact. Crucially, the Radon range-space content used in this process is estimated by a separate neural network with a combined objective. We demonstrate the effectiveness and generalizability of RN-SDEs through comprehensive experiments on several LACT scenarios based on two different datasets with distinct data distributions. In summary, RN-SDEs offer a robust and efficient solution for LACT reconstruction, achieving SOTA performance while maintaining computational efficiency. This work paves the way for future research to apply diffusion-based methods to solve various IR tasks and explore their applications in broader imaging contexts.
\\
\bibliographystyle{IEEEtran}
\bibliography{refs}

@article{withers2021x,
  title={X-ray computed tomography},
  author={Withers, Philip J and Bouman, Charles and Carmignato, Simone and Cnudde, Veerle and Grimaldi, David and Hagen, Charlotte K and Maire, Eric and Manley, Marena and Du Plessis, Anton and Stock, Stuart R},
  journal={Nature Reviews Methods Primers},
  volume={1},
  number={1},
  pages={18},
  year={2021},
  publisher={Nature Publishing Group UK London}
}

@article{ercius2015electron,
  title={Electron tomography: a three-dimensional analytic tool for hard and soft materials research},
  author={Ercius, Peter and Alaidi, Osama and Rames, Matthew J and Ren, Gang},
  journal={Advanced materials},
  volume={27},
  number={38},
  pages={5638--5663},
  year={2015},
  publisher={Wiley Online Library}
}

@article{muehllehner2006positron,
  title={Positron emission tomography},
  author={Muehllehner, Gerd and Karp, Joel S},
  journal={Physics in Medicine \& Biology},
  volume={51},
  number={13},
  pages={R117},
  year={2006},
  publisher={IOP Publishing}
}

@article{bracewell1956strip,
  title={Strip integration in radio astronomy},
  author={Bracewell, Ronald N},
  journal={Australian Journal of Physics},
  volume={9},
  number={2},
  pages={198--217},
  year={1956},
  publisher={CSIRO Publishing}
}

@inproceedings{sohl2015deep,
  title={Deep unsupervised learning using nonequilibrium thermodynamics},
  author={Sohl-Dickstein, Jascha and Weiss, Eric and Maheswaranathan, Niru and Ganguli, Surya},
  booktitle={International conference on machine learning},
  pages={2256--2265},
  year={2015},
  organization={PMLR}
}

@article{radon69u,
  title={U ber die Bestimmung von Funktionen durch ihre Integralwerte langs gewisser Mannigfaltigkeiten; Ber. vor Sachs},
  author={Radon, JH},
  journal={Akad. Wiss},
  year={1917},
  volume={69},
  pages={262}
}

@article{alberti2021learning,
  title={Learning the optimal Tikhonov regularizer for inverse problems},
  author={Alberti, Giovanni S and De Vito, Ernesto and Lassas, Matti and Ratti, Luca and Santacesaria, Matteo},
  journal={Advances in Neural Information Processing Systems},
  volume={34},
  pages={25205--25216},
  year={2021}
}

@article{danielyan2011bm3d,
  title={BM3D frames and variational image deblurring},
  author={Danielyan, Aram and Katkovnik, Vladimir and Egiazarian, Karen},
  journal={IEEE Transactions on image processing},
  volume={21},
  number={4},
  pages={1715--1728},
  year={2011},
  publisher={IEEE}
}

@article{barutcu2021limited,
  title={Limited-angle computed tomography with deep image and physics priors},
  author={Barutcu, Semih and Aslan, Selin and Katsaggelos, Aggelos K and G{\"u}rsoy, Do{\u{g}}a},
  journal={Scientific reports},
  volume={11},
  number={1},
  pages={17740},
  year={2021},
  publisher={Nature Publishing Group UK London}
}

@inproceedings{anirudh2018lose,
  title={Lose the views: Limited angle CT reconstruction via implicit sinogram completion},
  author={Anirudh, Rushil and Kim, Hyojin and Thiagarajan, Jayaraman J and Mohan, K Aditya and Champley, Kyle and Bremer, Timo},
  booktitle={Proceedings of the IEEE Conference on Computer Vision and Pattern Recognition},
  pages={6343--6352},
  year={2018}
}

@article{gupta2018cnn,
  title={CNN-based projected gradient descent for consistent CT image reconstruction},
  author={Gupta, Harshit and Jin, Kyong Hwan and Nguyen, Ha Q and McCann, Michael T and Unser, Michael},
  journal={IEEE transactions on medical imaging},
  volume={37},
  number={6},
  pages={1440--1453},
  year={2018},
  publisher={IEEE}
}

@article{han2018framing,
  title={Framing U-Net via deep convolutional framelets: Application to sparse-view CT},
  author={Han, Yoseob and Ye, Jong Chul},
  journal={IEEE transactions on medical imaging},
  volume={37},
  number={6},
  pages={1418--1429},
  year={2018},
  publisher={IEEE}
}

@article{jin2017deep,
  title={Deep convolutional neural network for inverse problems in imaging},
  author={Jin, Kyong Hwan and McCann, Michael T and Froustey, Emmanuel and Unser, Michael},
  journal={IEEE transactions on image processing},
  volume={26},
  number={9},
  pages={4509--4522},
  year={2017},
  publisher={IEEE}
}

@article{zhang2020artifact,
  title={Artifact removal using a hybrid-domain convolutional neural network for limited-angle computed tomography imaging},
  author={Zhang, Qiyang and Hu, Zhanli and Jiang, Changhui and Zheng, Hairong and Ge, Yongshuai and Liang, Dong},
  journal={Physics in Medicine \& Biology},
  volume={65},
  number={15},
  pages={155010},
  year={2020},
  publisher={IOP Publishing}
}

@article{zhang2021cd,
  title={CD-Net: Comprehensive domain network with spectral complementary for DECT sparse-view reconstruction},
  author={Zhang, Yikun and Lv, Tianling and Ge, Rongjun and Zhao, Qianlong and Hu, Dianlin and Zhang, Liu and Liu, Jin and Zhang, Yi and Liu, Qiegen and Zhao, Wei and others},
  journal={IEEE Transactions on Computational Imaging},
  volume={7},
  pages={436--447},
  year={2021},
  publisher={IEEE}
}

@article{goodfellow2014generative,
  title={Generative adversarial nets},
  author={Goodfellow, Ian and Pouget-Abadie, Jean and Mirza, Mehdi and Xu, Bing and Warde-Farley, David and Ozair, Sherjil and Courville, Aaron and Bengio, Yoshua},
  journal={Advances in neural information processing systems},
  volume={27},
  year={2014}
}

@article{kingma2013auto,
  title={Auto-encoding variational bayes},
  author={Kingma, Diederik P and Welling, Max},
  journal={arXiv preprint arXiv:1312.6114},
  year={2013}
}

@article{van2017neural,
  title={Neural discrete representation learning},
  author={Van Den Oord, Aaron and Vinyals, Oriol and others},
  journal={Advances in neural information processing systems},
  volume={30},
  year={2017}
}

@inproceedings{karras2019style,
  title={A style-based generator architecture for generative adversarial networks},
  author={Karras, Tero and Laine, Samuli and Aila, Timo},
  booktitle={Proceedings of the IEEE/CVF conference on computer vision and pattern recognition},
  pages={4401--4410},
  year={2019}
}

@article{ho2020denoising,
  title={Denoising diffusion probabilistic models},
  author={Ho, Jonathan and Jain, Ajay and Abbeel, Pieter},
  journal={Advances in neural information processing systems},
  volume={33},
  pages={6840--6851},
  year={2020}
}

@article{song2019generative,
  title={Generative modeling by estimating gradients of the data distribution},
  author={Song, Yang and Ermon, Stefano},
  journal={Advances in neural information processing systems},
  volume={32},
  year={2019}
}

@article{song2020score,
  title={Score-based generative modeling through stochastic differential equations},
  author={Song, Yang and Sohl-Dickstein, Jascha and Kingma, Diederik P and Kumar, Abhishek and Ermon, Stefano and Poole, Ben},
  journal={arXiv preprint arXiv:2011.13456},
  year={2020}
}

@inproceedings{wang2023zero,
  title={Zero-Shot Image Restoration Using Denoising Diffusion Null-Space Model},
  author={Wang, Yinhuai and Yu, Jiwen and Zhang, Jian},
  booktitle={The Eleventh International Conference on Learning Representations},
  year={2023}
}

@inproceedings{liu2023dolce,
  title={DOLCE: A model-based probabilistic diffusion framework for limited-angle ct reconstruction},
  author={Liu, Jiaming and Anirudh, Rushil and Thiagarajan, Jayaraman J and He, Stewart and Mohan, K Aditya and Kamilov, Ulugbek S and Kim, Hyojin},
  booktitle={Proceedings of the IEEE/CVF International Conference on Computer Vision},
  pages={10498--10508},
  year={2023}
}

@article{choi2021ilvr,
  title={Ilvr: Conditioning method for denoising diffusion probabilistic models},
  author={Choi, Jooyoung and Kim, Sungwon and Jeong, Yonghyun and Gwon, Youngjune and Yoon, Sungroh},
  journal={arXiv preprint arXiv:2108.02938},
  year={2021}
}

@article{dhariwal2021diffusion,
  title={Diffusion models beat gans on image synthesis},
  author={Dhariwal, Prafulla and Nichol, Alexander},
  journal={Advances in neural information processing systems},
  volume={34},
  pages={8780--8794},
  year={2021}
}

@article{kudo2013image,
  title={Image reconstruction for sparse-view CT and interior CT—introduction to compressed sensing and differentiated backprojection},
  author={Kudo, Hiroyuki and Suzuki, Taizo and Rashed, Essam A},
  journal={Quantitative imaging in medicine and surgery},
  volume={3},
  number={3},
  pages={147},
  year={2013},
  publisher={AME Publications}
}

@article{song2021solving,
  title={Solving inverse problems in medical imaging with score-based generative models},
  author={Song, Yang and Shen, Liyue and Xing, Lei and Ermon, Stefano},
  journal={arXiv preprint arXiv:2111.08005},
  year={2021}
}

@article{schwab2019deep,
  title={Deep null space learning for inverse problems: convergence analysis and rates},
  author={Schwab, Johannes and Antholzer, Stephan and Haltmeier, Markus},
  journal={Inverse Problems},
  volume={35},
  number={2},
  pages={025008},
  year={2019},
  publisher={IOP Publishing}
}

@inproceedings{chwialkowski2016kernel,
  title={A kernel test of goodness of fit},
  author={Chwialkowski, Kacper and Strathmann, Heiko and Gretton, Arthur},
  booktitle={International conference on machine learning},
  pages={2606--2615},
  year={2016},
  organization={PMLR}
}

@inproceedings{liu2016kernelized,
  title={A kernelized Stein discrepancy for goodness-of-fit tests},
  author={Liu, Qiang and Lee, Jason and Jordan, Michael},
  booktitle={International conference on machine learning},
  pages={276--284},
  year={2016},
  organization={PMLR}
}

@inproceedings{chung2023diffusion,
  title={Diffusion Posterior Sampling for General Noisy Inverse Problems},
  author={Chung, Hyungjin and Kim, Jeongsol and Mccann, Michael T and Klasky, Marc L and Ye, Jong Chul},
  booktitle={The Eleventh International Conference on Learning Representations, ICLR 2023},
  year={2023},
  organization={The International Conference on Learning Representations}
}

@article{chung2022improving,
  title={Improving diffusion models for inverse problems using manifold constraints},
  author={Chung, Hyungjin and Sim, Byeongsu and Ryu, Dohoon and Ye, Jong Chul},
  journal={Advances in Neural Information Processing Systems},
  volume={35},
  pages={25683--25696},
  year={2022}
}

@article{luo2023image,
  title={Image restoration with mean-reverting stochastic differential equations},
  author={Luo, Ziwei and Gustafsson, Fredrik K and Zhao, Zheng and Sj{\"o}lund, Jens and Sch{\"o}n, Thomas B},
  journal={arXiv preprint arXiv:2301.11699},
  year={2023}
}

@article{saharia2022image,
  title={Image super-resolution via iterative refinement},
  author={Saharia, Chitwan and Ho, Jonathan and Chan, William and Salimans, Tim and Fleet, David J and Norouzi, Mohammad},
  journal={IEEE transactions on pattern analysis and machine intelligence},
  volume={45},
  number={4},
  pages={4713--4726},
  year={2022},
  publisher={IEEE}
}

@article{li2022srdiff,
  title={Srdiff: Single image super-resolution with diffusion probabilistic models},
  author={Li, Haoying and Yang, Yifan and Chang, Meng and Chen, Shiqi and Feng, Huajun and Xu, Zhihai and Li, Qi and Chen, Yueting},
  journal={Neurocomputing},
  volume={479},
  pages={47--59},
  year={2022},
  publisher={Elsevier}
}

@article{kawar2021snips,
  title={Snips: Solving noisy inverse problems stochastically},
  author={Kawar, Bahjat and Vaksman, Gregory and Elad, Michael},
  journal={Advances in Neural Information Processing Systems},
  volume={34},
  pages={21757--21769},
  year={2021}
}

@inproceedings{penrose1955generalized,
  title={A generalized inverse for matrices},
  author={Penrose, Roger},
  booktitle={Mathematical proceedings of the Cambridge philosophical society},
  volume={51},
  pages={406--413},
  year={1955},
  organization={Cambridge University Press}
}

@article{dudgeon1983multidimensional,
  title={Multidimensional digital signal processing},
  author={Dudgeon, Dan E},
  journal={Engewood Cliffs},
  year={1983},
  publisher={Prentice-hall}
}

@article{croitoru2023diffusion,
  title={Diffusion models in vision: A survey},
  author={Croitoru, Florinel-Alin and Hondru, Vlad and Ionescu, Radu Tudor and Shah, Mubarak},
  journal={IEEE Transactions on Pattern Analysis and Machine Intelligence},
  year={2023},
  publisher={IEEE}
}

@article{yang2023diffusion,
  title={Diffusion models: A comprehensive survey of methods and applications},
  author={Yang, Ling and Zhang, Zhilong and Song, Yang and Hong, Shenda and Xu, Runsheng and Zhao, Yue and Zhang, Wentao and Cui, Bin and Yang, Ming-Hsuan},
  journal={ACM Computing Surveys},
  volume={56},
  number={4},
  pages={1--39},
  year={2023},
  publisher={ACM New York, NY, USA}
}

@inproceedings{nichol2021improved,
  title={Improved denoising diffusion probabilistic models},
  author={Nichol, Alexander Quinn and Dhariwal, Prafulla},
  booktitle={International conference on machine learning},
  pages={8162--8171},
  year={2021},
  organization={PMLR}
}

@article{song2020improved,
  title={Improved techniques for training score-based generative models},
  author={Song, Yang and Ermon, Stefano},
  journal={Advances in neural information processing systems},
  volume={33},
  pages={12438--12448},
  year={2020}
}

@article{song2021maximum,
  title={Maximum likelihood training of score-based diffusion models},
  author={Song, Yang and Durkan, Conor and Murray, Iain and Ermon, Stefano},
  journal={Advances in neural information processing systems},
  volume={34},
  pages={1415--1428},
  year={2021}
}

@article{elliott1985reverse,
  title={Reverse time diffusions},
  author={Elliott, Robert J and Anderson, Brian DO},
  journal={Stochastic processes and their applications},
  volume={19},
  number={2},
  pages={327--339},
  year={1985},
  publisher={Elsevier}
}

@article{hyvarinen2005estimation,
  title={Estimation of non-normalized statistical models by score matching.},
  author={Hyv{\"a}rinen, Aapo and Dayan, Peter},
  journal={Journal of Machine Learning Research},
  volume={6},
  number={4},
  year={2005}
}

@article{vincent2011connection,
  title={A connection between score matching and denoising autoencoders},
  author={Vincent, Pascal},
  journal={Neural computation},
  volume={23},
  number={7},
  pages={1661--1674},
  year={2011},
  publisher={MIT Press}
}

@inproceedings{song2020sliced,
  title={Sliced score matching: A scalable approach to density and score estimation},
  author={Song, Yang and Garg, Sahaj and Shi, Jiaxin and Ermon, Stefano},
  booktitle={Uncertainty in Artificial Intelligence},
  pages={574--584},
  year={2020},
  organization={PMLR}
}

@inproceedings{chen2022simple,
  title={Simple baselines for image restoration},
  author={Chen, Liangyu and Chu, Xiaojie and Zhang, Xiangyu and Sun, Jian},
  booktitle={European conference on computer vision},
  pages={17--33},
  year={2022},
  organization={Springer}
}

@inproceedings{blau2018perception,
  title={The perception-distortion tradeoff},
  author={Blau, Yochai and Michaeli, Tomer},
  booktitle={Proceedings of the IEEE conference on computer vision and pattern recognition},
  pages={6228--6237},
  year={2018}
}

@article{carignano2024local,
  title={Local volume concentration, packing domains, and scaling properties of chromatin},
  author={Carignano, Marcelo A and Kr{\"o}ger, Martin and Almassalha, Luay M and Agrawal, Vasundhara and Li, Wing Shun and Pujadas-Liwag, Emily M and Nap, Rikkert J and Backman, Vadim and Szleifer, Igal},
  journal={eLife},
  volume={13},
  pages={RP97604},
  year={2024},
  publisher={eLife Sciences Publications Limited}
}

@article{heller2019kits19,
  title={The kits19 challenge data: 300 kidney tumor cases with clinical context, ct semantic segmentations, and surgical outcomes},
  author={Heller, Nicholas and Sathianathen, Niranjan and Kalapara, Arveen and Walczak, Edward and Moore, Keenan and Kaluzniak, Heather and Rosenberg, Joel and Blake, Paul and Rengel, Zachary and Oestreich, Makinna and others},
  journal={arXiv preprint arXiv:1904.00445},
  year={2019}
}

@article{heller2021state,
  title={The state of the art in kidney and kidney tumor segmentation in contrast-enhanced CT imaging: Results of the KiTS19 challenge},
  author={Heller, Nicholas and Isensee, Fabian and Maier-Hein, Klaus H and Hou, Xiaoshuai and Xie, Chunmei and Li, Fengyi and Nan, Yang and Mu, Guangrui and Lin, Zhiyong and Han, Miofei and others},
  journal={Medical image analysis},
  volume={67},
  pages={101821},
  year={2021},
  publisher={Elsevier}
}

@article{chambolle2011first,
  title={A first-order primal-dual algorithm for convex problems with applications to imaging},
  author={Chambolle, Antonin and Pock, Thomas},
  journal={Journal of mathematical imaging and vision},
  volume={40},
  pages={120--145},
  year={2011},
  publisher={Springer}
}

@article{gursoy2014tomopy,
  title={TomoPy: a framework for the analysis of synchrotron tomographic data},
  author={G{\"u}rsoy, Doga and De Carlo, Francesco and Xiao, Xianghui and Jacobsen, Chris},
  journal={Journal of synchrotron radiation},
  volume={21},
  number={5},
  pages={1188--1193},
  year={2014},
  publisher={International Union of Crystallography}
}

@inproceedings{whang2022deblurring,
  title={Deblurring via stochastic refinement},
  author={Whang, Jay and Delbracio, Mauricio and Talebi, Hossein and Saharia, Chitwan and Dimakis, Alexandros G and Milanfar, Peyman},
  booktitle={Proceedings of the IEEE/CVF Conference on Computer Vision and Pattern Recognition},
  pages={16293--16303},
  year={2022}
}

@inproceedings{zhang2018unreasonable,
  title={The unreasonable effectiveness of deep features as a perceptual metric},
  author={Zhang, Richard and Isola, Phillip and Efros, Alexei A and Shechtman, Eli and Wang, Oliver},
  booktitle={Proceedings of the IEEE conference on computer vision and pattern recognition},
  pages={586--595},
  year={2018}
}

@inproceedings{luo2023refusion,
  title={Refusion: Enabling Large-Size Realistic Image Restoration with Latent-Space Diffusion Models},
  author={Luo, Ziwei and Gustafsson, Fredrik K and Zhao, Zheng and Sj{\"o}lund, Jens and Sch{\"o}n, Thomas B},
  booktitle={Proceedings of the IEEE/CVF Conference on Computer Vision and Pattern Recognition Workshops},
  pages={1680--1691},
  year={2023}
}

\appendix
\section{Appendix}

\subsection{Generalize DDPMs and SGMs to SDEs}\label{Appendix: DDPMs SGMs to SDEs}
Recall the one-step DDPM~\cite{ho2020denoising} forward diffusion process, which gradually transforms the data distribution $q(\mathbf{x}_0)$ into noise.
$$
\mathbf{x}_t=\sqrt{1-\beta_t} \,\mathbf{x}_{t-1}+\sqrt{\beta_t} \,\mathcal{N}(0, \mathbf{I}),\quad \forall t \in\{1, \ldots, T\}.
$$
In SDEs, the diffusion process is considered to be continuous, thus becoming the solution of an SDE. In other words, we will take infinite steps and each step will be infinitely small. Therefore, we have:
$$
\begin{aligned}
\mathbf{x}_t & =\sqrt{1-\beta_t}\, \mathbf{x}_{t-1}+\sqrt{\beta_t} \,\mathcal{N}(\mathbf{0}, \mathbf{I})\\
& \approx \mathbf{x}_{t-1}-\frac{\beta(t) \Delta t}{2} \mathbf{x}_{t-1}+\sqrt{\beta(t) \Delta t}\, \mathcal{N}(\mathbf{0}, \mathbf{I}),
\end{aligned}
$$
where the last approximation is obtained using Taylor expansion. This iterative update corresponds to a certain solution or a certain discretization of an SDE:
$$
\mathrm{d} \mathbf{x}=\underbrace{-\frac{1}{2} \beta(t) \mathbf{x}\,\mathrm{d} t}_{\text {drift term }} +\underbrace{\sqrt{\beta(t)} \,\mathrm{d} \mathbf{w}}_{\text {diffusion term}}.
$$
Similarly, one can also easily start from the one-step SGM~\cite{song2019generative} forward diffusion process:
$$
\mathbf{x}_t=\mathbf{x}_{t-1}+\sqrt{\sigma_t^2-\sigma_{t-1}^2} \,\mathcal{N}(0, \mathbf{I}),\quad \forall t \in\{1, \ldots, T\},
$$
to its corresponding SDE version:
$$
\mathrm{d} \mathbf{x}=\sqrt{\frac{\mathrm{d}\left[\sigma(t)^2\right]}{\mathrm{d} t}} \mathrm{~d} \mathbf{w}.
$$

\subsection{RNSD With Maximum Likelihood Objective}\label{Appendix: RNSD for RN-SDE}
The training process in this paper follows a manner similar to conventional score matching. However, the objective differs as it aims to identify an $\mathbf{x}_{t-1}$ that maximizes the log-likelihood, $\log p(\mathbf{x}_{t-1} | \mathbf{x}_t, \mathbf{x}_{0})$. A theoretical optimal solution, $\mathbf{x}_{t-1}^*$, is shown by~\cite{luo2023image} as:
$$
\begin{aligned}
\mathbf{x}_{t-1}^*= & \;\frac{1-\mathrm{e}^{-2 \bar{\zeta}_{t-1}}}{1-\mathrm{e}^{-2 \bar{\zeta}_t}} \mathrm{e}^{-\zeta_t^{\prime}}\left(\mathbf{x}_t-\bmu\right) \\
& +\frac{1-\mathrm{e}^{-2 \zeta_t^{\prime}}}{1-\mathrm{e}^{-2 \bar{\zeta}_t}} \mathrm{e}^{-\bar{\zeta}_{t-1}}\left(\mathbf{x}_0-\bmu\right)+\bmu.
\end{aligned}
$$
At the same time, recall the reverse MR-SDE described in Eq.~\ref{reverse mr sde}. By replacing $\mathrm{d} \mathbf{x}$ with $\mathbf{x}_{t}-\mathbf{x}_{t-1}$ and rearranging, we have:
$$
\begin{aligned}
   \mathbf{x}_{t-1} = \mathbf{x}_t - \bigg[ \zeta_t (\bmu - \mathbf{x}_t) - \sigma_t^2\nabla_\mathbf{x} \log p(\mathbf{x}_t)\bigg] \mathrm{d} t+ \sigma_t \mathrm{~d} \widehat{\mathbf{w}}.
\end{aligned}
$$
Bringing $\mathbf{x}_{t-1}^*$ into the above equation and simplifying, we arrive at two important derivations, the first of which corresponds to Eq.~\ref{opt score}, which states that the objective of maximum likelihood is equivalent to estimating the optimal score, $\nabla_\mathbf{x} \log p(\mathbf{x}_t | \mathbf{x}_0)^*$:
$$
\nabla_\mathbf{x} \log p(\mathbf{x}_t | \mathbf{x}_0)^* =  \frac{g_t\mathbf{x}_t+h_t\mathbf{x}_0-(h_t+g_t)\bmu}{\sigma_t^2 \mathrm{d} t}.
$$
Based on this, we can obtain the following training objective:
$$
\mathbb{E}_{t \in \mathcal{U}}  \mathbb{E}_{p(\mathbf{x}_t)}\left[\left\|\nabla_{\mathbf{x}} \log p(\mathbf{x}_t|\mathbf{x}_0)^*-\hat{\mathbf{s}}_\phi(\mathbf{x}_t, \bmu, t)\right\|\right],
$$
that is to minimize the $l_2$ distance between the theoretical optimal score and the network's estimation.

Another useful derivation corresponds to Eq.~\ref{clean inter-states} in the paper, which provides a clean intermediate state $\mathbf{x}_{0|t}$ that is easy to process. In this work, we apply RNSD-based rectification to this estimate, yielding a corrected image ${\mathbf{x}}_{0|t}^{\prime}$. We use ${\mathbf{x}}_{0|t}^{\prime}$ to update the score in Eq.~\ref{opt score} by replacing $\mathbf{x}_0$ with ${\mathbf{x}}_{0|t}^{\prime}$ and integrate the updated score into the reverse MR-SDE to complete one refinement iteration. This can be summarized as:
$$
\begin{aligned}
  \mathbf{x}_{t-1} &=  h_t{\mathbf{x}}_{0|t}^{\prime} + (g_t + \zeta_t\mathrm{d}t)\mathbf{x}_t \\
  & \quad\quad - (h_t+g_t+\zeta_t\mathrm{d}t)\bmu + \sigma_t\sqrt{\mathrm{d}t}\,\bepsilon_t, \quad \bepsilon_t \sim \mathcal{N}(0, \mathbf{I}).
\end{aligned}
$$
By replacing $\mathbf{x}_{t-1}$ with $\mathbf{x}_{t}-\mathrm{d}\mathbf{x}$ and rearranging, we reached the reverse RN-SDE expression in Eq.~\ref{xt-1}.

\subsection{Difficulties of Computing Radon Pseudo-Inverse}\label{Appendix: complexity of A+}
To elucidate the computational dilemma associated with the Radon pseudo-inverse, $\mathbf{A}^{\dagger}$, let us consider an input of size $\mathbf{x}\in \mathbb{R}^{M}$. The Radon transform of this input is represented by a matrix  $\mathbf{A}\in\mathbb{R}^{M\times N}$, which projects $\mathbf{x}$ into the sinogram domain, i.e., $\mathbf{y}\in\mathbb{R}^{N}$. Then, a standard approach of computing $\mathbf{A}^{\dagger}$ is to use Singular Value Decomposition (SVD),

\begin{eqnarray}
&\mathbf{A} = \mathbf{V} \boldsymbol{\Sigma} \mathbf{U}^{\top},  
\quad \mathbf{A}^{\dagger} = \mathbf{U} \boldsymbol{\Sigma}^{\dagger} \mathbf{V}^{\top}, \qquad\qquad\qquad\qquad & \hfill \nonumber \\
&\boldsymbol{\Sigma} = \operatorname{diag}\left\{s_1, s_2, \cdots, s_k\right\},  \quad \boldsymbol{\Sigma}^{\dagger}_i = 
\begin{cases} \frac{1}{s_i}, & s_i \neq 0, \\ 
0, & s_i = 0 
\end{cases}, &
\end{eqnarray}
with $\mathbf{A},\boldsymbol{\Sigma} \in \mathbb{R}^{N \times M}$, $\mathbf{A}^{\dagger},\boldsymbol{\Sigma}^{\dagger} \in \mathbb{R}^{M \times N}$, $\mathbf{U} \in \mathbb{R}^{M \times M}$, $\mathbf{V} \in \mathbb{R}^{N \times N}$.
The computational complexity of performing SVD on $\mathbf{A}$ is $O(N^2M+M^3$) (assume $N>M$), which makes this calculation impractical due to the significant computational resources required. For example, in the context of this paper (C4KC-KiTS), $M=512\times512$, while $N=720\times512$.

\subsection{Implementation of Learnable Radon Back-Projection}\label{Appendix: Pseudo_inverse}
For any arbitrary 2D object $f(x, y)$, we consider the tomographic measurement process at a given angle $\theta$ to consist of a set of line integrals and the sinogram is the collection of such scans over a set of angles. Therefore, the value of any position $r$ in the image, where the point $(x,y)$ will be projected at angle $\theta$ can be measured by:
$$
x \cos \theta + y \sin \theta = r,
$$
and the corresponding Radon transform can be written as: 
$$
\mathrm{q}_{\theta}(r) = \int_{-\infty}^{\infty} \int_{-\infty}^{\infty} f(x, y) \delta(x \cos \theta + y \sin \theta - r) \, \mathrm{d}x \, \mathrm{d}y,
$$
where $\delta(\cdot)$ denotes the Dirac function and $p_{\theta}(r)$ represents the Radon projections, which is the input to the Learnable Radon Back-Projection. To reverse this process, similar to FBP, we use a learnable Ramp filter to filter $p_{\theta}(r)$. This filtering is typically performed by transforming the frequency response of the Ramp filter with a simple linear layer. When this transformation is set as the identity function, the learnable back-projection simplifies to standard FBP (recommended for some simple cases). For simplicity, we denote the learnable Ramp filter as $f_\eta$. The filtered sinogram, $\mathrm{q}^*_{\theta}(r)$ is given by:
$$
\mathrm{q}^*_{\theta}(r) = \mathcal{F}^{-1} \left\{ |f_\eta| \cdot \mathcal{F} \{ \mathrm{q}_{\theta}(r)  \} \right\},
$$
where $f_\eta$ is implemented as a simple linear layer to refine the frequency response of the Ramp filter, and $\mathcal{F}$ represents the Fourier transform. This filtered sinogram can then be projected back to the image domain through:
$$
\mathrm{f}(x, y) = \int_0^{\pi} \mathrm{q}_{\theta}(r)  \, \mathrm{d} \theta,
$$
where $\mathrm{d} \theta$ is the angular spacing between the projections.

\section{Biography Section}




\end{document}